\documentclass[showpacs, preprintnumbers, nofootinbib, aps, prd,
               superscriptaddress,10pt, showkeys, notitlepage,twocolumn]{revtex4-1}

\usepackage{graphicx,amssymb,amsmath,amsthm,amsfonts,epsfig}

\usepackage[linktocpage]{hyperref}
\usepackage[usenames,dvipsnames]{color}
\usepackage{epstopdf}
\usepackage{aas_macros}
\usepackage{pifont}
\usepackage[normalem]{ulem}
\usepackage{tabularx}
\definecolor{darkred}{rgb}{0.5,0,0}
\definecolor{darkgreen}{rgb}{0,0.5,0}
\definecolor{darkblue}{rgb}{0,0,0.5}
\definecolor{prussian}{rgb}{0.0, 0.19, 0.33}
\definecolor{richelectricblue}{rgb}{0.03, 0.57, 0.82}
\definecolor{teal}{rgb}{0.0, 0.5, 0.5}
\definecolor{mediumseagreen}{rgb}{0.24, 0.7, 0.44}
\definecolor{lust}{rgb}{0.9, 0.13, 0.13}
\definecolor{ballblue}{rgb}{0.13, 0.67, 0.8}
\definecolor{darkcyan}{rgb}{0.0, 0.55, 0.55}
\definecolor{mountainmeadow}{rgb}{0.19, 0.73, 0.56}
\definecolor{palecarmine}{rgb}{0.69, 0.25, 0.21}
\definecolor{richcarmine}{rgb}{0.84, 0.0, 0.25}
\definecolor{tangelo}{rgb}{0.98, 0.3, 0.0}
\definecolor{venetian}{rgb}{0.784,0.031,0.082}
\definecolor{bdfrance}{rgb}{0.192,0.549,0.906}

\hypersetup{colorlinks=true, citecolor=venetian,
            linkcolor=bdfrance, urlcolor=lust}
\usepackage{amsmath,amssymb}
\usepackage{tensor}
\usepackage{mathtools}
\usepackage{amsbsy}
\usepackage{bm}
\usepackage{float}

\newcommand{\be}{\begin{equation}}
\newcommand{\ee}{\end{equation}}
\newcommand{\bea}{\begin{eqnarray}}
\newcommand{\eea}{\end{eqnarray}}
\newcommand{\nn}{\nonumber}

\newcommand{\p}{\prime}
\newcommand{\pp}{\prime\prime}

\newcommand{\rph}{r_{\rm ph}}
\newcommand{\rmax}{r_{\rm max}}

\begin{document}

\title{Eikonal quasinormal modes of black holes beyond General Relativity}

\author{Kostas  Glampedakis}
\email{kostas@um.es}
\affiliation{Departamento de F\'isica, Universidad de Murcia, Murcia E-30100, Spain}
\affiliation{Theoretical Astrophysics, University of T\"ubingen, Auf der Morgenstelle 10, T\"ubingen, D-72076, Germany}

\author{Hector O. Silva}
\email{hector.okadadasilva@montana.edu}
\affiliation{eXtreme Gravity Institute, Department of Physics, Montana State University, Bozeman, Montana 59717, USA}

\date{{\today}}

\begin{abstract}

Much of our physical intuition about black hole quasinormal modes in general relativity comes from the
eikonal/geometric optics approximation. According to the well-established eikonal model, the fundamental quasinormal
mode represents wavepackets orbiting in the vicinity of the black hole's geodesic photon ring, slowly peeling off towards the
event horizon and infinity. Besides its strength as a `visualisation' tool, the eikonal approximation also provides a simple
quantitative method for calculating the mode frequency, in close agreement with rigorous numerical results.
In this paper we move away from Einstein's theory and its garden-variety black holes and go on to consider spherically
symmetric black holes in modified theories of gravity through the lens of the eikonal approximation. The quasinormal modes of such
black holes are typically described by a set of coupled wave equations for the various field degrees of freedom.
Considering a general, theory-agnostic, system of two equations for two perturbed fields, we derive eikonal formulae for the complex
fundamental quasinormal mode frequency. In addition we show that the eikonal modes can be related to the extremum of an effective
potential and its associated `photon ring'. As an application of our results we consider a specific example of a modified theory of gravity with
known black hole quasinormal modes and find that these are well approximated by the eikonal formulae.

\end{abstract}

\maketitle

\section{Introduction}
\label{sec:intro}

A full century after the conception of general relativity (GR) the direct observation of gravitational waves (GWs) from merging black holes
by the advanced LIGO-Virgo network of detectors~\cite{Abbott:2016blz, TheLIGOScientific:2016pea, Abbott:2016nmj, Abbott:2017vtc, GW170814}
has finally opened the door to tests of relativistic gravity in the truly nonlinear strong field regime. Among the  prime objectives of present and
near-future GW astronomy is the materialisation of `black hole spectroscopy', that is, the observation of the so-called ringdown signal at the very
end of the merger and the extraction/identification of the final black hole's quasinormal modes (QNMs). The power of this method, much like its
atomic physics kin, lies in the fact that GR predicts a unique spectrum of complex QNM frequencies for a given black hole mass and spin and therefore
the simultaneous observation of more than two QNMs should, in principle, allow the Kerr hypothesis to be tested \cite{Detweiler:1980gk,Dreyer:2003bv,Berti:2005ys}.

Any programme aiming at probing the true nature of black holes should allow for deviations from GR's Kerr spacetime as well as for
theoretical input from alternative to GR theories of gravity. Perhaps the simplest `beyond-Kerr' strategy  is to use parametrised schemes (with the parameters controlling
the deformation away from GR) both for the black hole's spacetime metric~\cite{Johannsen:2011dh, Johannsen2013PhRvD,Konoplya:2016jvv,Cardoso:2014rha}
and the associated QNMs \cite{Glampedakis:2017, Cardoso_etal2019,McManus_etal2019} without the need to commit to any particular theory
of gravity.  The main drawback of this approach is that it may constitute nothing more than a null test of the Kerr metric, in the sense that the
deformations may not actually map onto any specific gravity theory. The alternative, more rigorous (and far more laborious) strategy is the
theoretical calculation of black hole spacetimes and their GW signature on a case-to-case basis within the zoo of modified gravity theories.
Not surprisingly, this second approach is much more difficult to implement and as a result QNMs of non-GR black holes have been computed
only for a handful of cases, usually under the assumption of spherical
symmetry, e.g.~\cite{molina2010, Kobayashi:2012kh, Kobayashi:2014wsa,Blazquez-Salcedo:2016enn, Blazquez-Salcedo:2016enn,Blazquez-Salcedo:2017txk, Brito:2018hjh, Tattersall_2018a, Tattersall_etal2018}
(for a comprehensive review and further references see~\cite{Bertietal2015}).

In this paper we study QNMs of black holes beyond GR by combining elements of the two aforementioned approaches.
We assume \emph{spherically symmetric} black holes in which case, after separating out the angular dependence,  the wave dynamics
and QNMs of the system are described by a set of radial wave equations for the perturbed spacetime metric (the tensorial field)
and the other fields that are generically present in modified theories of gravity. In some instances one of the two symmetry sectors of the tensorial field
(axial/odd of polar/even modes) couples to the other fields while the remaining one is described by the same Regge-Wheeler or Zerilli equation
as in GR. Focusing, for obvious reasons, on the \emph{coupled} case, we adopt a largely theory-agnostic approach and postulate
a pair of wave equations for the tensorial and the extra `scalar' field with parametrised potentials.
Although far from representing the most general situation~\cite{Tattersall_etal2018}, our parametrised model provides a useful benchmark
for describing perturbed non-GR black holes; in addition it has the merit of including as a special case at least a pair of modified theories of gravity,
namely, dynamical Chern-Simons gravity~\cite{molina2010} and the sixth-order Proca theory~\cite{Heisenberg:2017hwb,Tattersall_etal2018}.

Our main main tool -- and second main simplification -- is the use of the \emph{eikonal} (or geometric optics) approximation
for obtaining QNM solutions from the wave equations. The eikonal approximation  has a long and successful history in the study of
Schwarzschild and Kerr black holes, dating back to the early 1970s~\cite{Press:1971wr, Goebel:1972}
(see~\cite{Kokkotas:1999bd, Berti:2009kk} for reviews on the subject). According to the established eikonal picture, the fundamental QNM
can be visualised as a wavepacket localised in the radial direction at the peak of the wave potential;  in this approximation the peak itself coincides
with the location of the photon ring of null geodesics. The real part of the QNM frequency is found to be an integer multiple of the orbital angular frequency
at the photon ring. Similarly, the decay rate (imaginary part) of the QNM is related to the Lyapunov exponent of the unstable null orbits at the photon ring
radius~\cite{Ferrari:1984zz, Mashhoon:1985cya,Cardoso:2008bp}. The same intuitive eikonal model has been used to establish a
connection between the $\ell > |m|$ QNMs of Kerr black holes (where $\ell, m$ are the usual spherical harmonic integers) and the nonequatorial
spherical photon orbits~\cite{Dolan:2010wr, Yang:2012he}. More recently, and inspired by the QNM-photon ring relation,
an eikonal post-Kerr parametrised scheme was developed as a model for the fundamental $\ell=|m|$ mode of non-GR black holes~\cite{Glampedakis:2017}.
Although this model can describe the astrophysically more interesting case of rotating black holes it also has the drawback that it does not
account for any extra field degrees of freedom. On the same topic of non-GR black holes it should be noted that some recent work has criticised
the validity of the connection between QNMs and photon geodesics~\cite{Khanna:2016yow,Konoplya:2017wot}, although the models and arguments
used in these papers are far from being conclusive.

The purpose of the eikonal calculation presented in this paper is therefore twofold. First, by including a coupling to an extra field,
we develop a parametrised eikonal QNM model that surpasses in rigorousness that of Ref.~\cite{Glampedakis:2017}
(although in doing so we restrict ourselves to the less physically interesting case of  spherical symmetry). Second, we examine to what
extent these novel eikonal formulae preserve some connection between the fundamental QNM and photon geodesics.
Our implementation of the eikonal approximation differs in one key aspect with respect to what is
usually done in the context of GR black holes. In this latter case, the eikonal approximation consists in
taking the angular limit $\ell \gg 1$  in the Bohr-Sommerfeld formula originating from the WKB-approximated radial
wave equations (for example, see \cite{Berti:2014bla}). As no WKB approximation appears to exist (to the best of our knowledge)
for a system of coupled wave equations, we are forced to take the eikonal limit of the equations themselves both in the radial and angular
directions. In essence, our approach is very similar to that of Ref.~\cite{Dolan:2010wr} for Kerr black holes, and this will become
apparent in the discussion of the following section.

The remainder of this paper is organised as follows. In Section~\ref{sec:wave_eik}, and before embarking on
our analysis of coupled wave equations, we study the eikonal limit of a single wave equation, albeit a generalised one that allows
for deviations from GR. Section~\ref{sec:nonGR} contains the main calculation of this paper, namely, the derivation of eikonal formulae
for non-GR QNMs described by a pair of coupled wave equations in the context of modified theories of gravity with an extra scalar field degree of
freedom besides the standard tensorial one. A summary of these results can be found in Section~\ref{sec:summary}.
Section~\ref{sec:CS} provides an application of our eikonal formulae for the case of QNMs
of Schwarzschild black holes in dynamical Chern-Simons gravity. Our concluding remarks can be found in Section~\ref{sec:conclusions}
while the two appendices contain some additional material on photon geodesics and their correspondence to the eikonal limit.

Throughout this paper we adopt geometric units $G=c=1$ and assume an $ e^{-i\omega t}$ time dependence for the perturbed fields.
We use a prime to denote a $d/dr$ radial derivative. For any function $f(r)$ we use abbreviations  $f_{\rm ph} = f(\rph), f_{\rm max} = f(\rmax)$, etc.


\section{Generalised wave equation in a spherical black hole spacetime}
\label{sec:wave_eik}

\subsection{Eikonal approximation: leading order}

In order to set the stage for calculating QNMs of non-GR black holes in the eikonal approximation we first consider the case
of a single (radial) wave equation describing perturbations of a field $\psi$ in a background spherical spacetime.
We assume an equation,
\be
\frac{d^2 \psi}{dx^2} + \left ( \omega^2 - U  \right ) \psi = 0,
\label{gen_wave}
\ee
where $x(r)$ is a suitably defined tortoise coordinate which maps the black hole horizon (located at $r=r_{\rm H}$) and infinity ($r=\infty$)
onto $x= -\infty$ and $x=+\infty$ respectively. The wave potential is assumed to be of the form\footnote{Single-field wave equations of the
form (\ref{gen_wave})-(\ref{UsinglePot}) can appear in extensions of GR like the higher-dimensional Einstein-Gauss-Bonnet
theory~\cite{Konoplya:2017wot}.},
\be
U = f(r) \left [\, \frac{\ell(\ell+1)}{r^2} \alpha(r) - \frac{6M}{r^3} \zeta (r) \, \right ],
\label{UsinglePot}
\ee
where $\ell$ is the familiar angular integer multipole and $f(r)$ a function with asymptotic behaviours $f(r\to r_{\rm H}) \to 0$ and $f(r \to \infty)\to 1$.
The functions $\alpha (r), \zeta (r)$ are assumed to carry no $\ell$-dependence but are otherwise unspecified. Crucially, we require
$U$ to be `black hole-like', that is, with a single peak and $U(x\to \pm \infty) \to 0 $.

According to the eikonal/geometric optics prescription  we look for wave solutions of (\ref{gen_wave}) of the form,
\be
\psi (x) = A(x) e^{iS(x)/\epsilon},
\ee
where $\epsilon$ is the customary bookkeeping parameter. Then,
\begin{eqnarray}
\frac{d^2 \psi}{dx} = e^{iS/\epsilon} \left [\, A_{,xx} + \frac{i}{\epsilon} \left (\,  2S_{,x} A_{,x} + S_{,xx} A \,\right ) - \frac{(S_{,x})^2}{\epsilon^2} A   \, \right ],
\nn \\
\end{eqnarray}
and the wave equation becomes,
\begin{align}
    A_{,xx} &+ \frac{i}{\epsilon} \left (\,  2S_{,x} A_{,x} + S_{,xx} A \,\right )
 + \Big [ \omega^2  - \frac{(S_{,x})^2}{\epsilon^2}
 \nn \\
 & - f\left (\, \frac{\ell(\ell+1)}{r^2} \alpha - \frac{6M}{r^3} \zeta \, \right ) \Big ] A = 0.
 \label{full_eik}
\end{align}
The double limit $\epsilon \ll 1$ and $\ell \gg 1$ enforces the eikonal limit in the radial and angular directions
and leads to the following leading-order equation,
\be
 - \frac{(S_{,x})^2}{\epsilon^2} + \omega^2 - \ell^2 \frac{f \alpha}{r^2}  = 0 + {\cal O}(\ell, \epsilon^{-1}).
 \label{lead_eik}
\ee
This expression, reminiscent of a radial Hamilton-Jacobi equation, only makes sense if the two expansion parameters balance,
$\epsilon \ell = {\cal O}(1)$. The physical picture behind this balance is that of wave packets equally localised in the radial and angular directions.
The same reasoning dictates a scaling $\omega = {\cal O} (\ell)$.

For a QNM wave solution we need to impose the boundary conditions,
\be
S (x\to  \pm \infty) = \pm \omega x.
\ee
Following a reasoning similar to the eikonal analysis of Ref. \cite{Dolan:2010wr},
we expect the phase function $S$ to switch from outgoing ($x >0$) to ingoing ($x<0$) at the location of the potential peak (located at $x\approx 0$);
in other words $S$ should have a minimum at that radius.

If we define $\tilde{U}$ as
\be
U(r) \approx  \ell^2   \frac{f \alpha}{r^2} \equiv \ell^2 \tilde{U} (r), \qquad (\ell \gg 1),
\ee
the potential peak in the eikonal limit, $r=r_0$, should be
\be
\tilde{U}^\p_0 = \left (  \frac{f \alpha}{r^2} \right )^\p_0 = 0,
\label{dU0eq}
\ee
where a prime stands for a derivative with respect to $r$.
In fact, the association of $S_{,x} = 0 $ with the potential peak can be enforced rather than being merely assumed
by taking the $x$-derivative of (\ref{lead_eik}):
\be
 \frac{2}{\epsilon^2} S_{,x} S_{,xx} = -  \ell^2 \frac{dr}{dx} \tilde{U}^\p ~\Rightarrow ~  S_{,x} (r_0) = 0.
\ee
The evaluation of (\ref{lead_eik}) at the potential peak $r=r_0$ singles out the black hole's fundamental QNM frequency (at leading
eikonal order). We obtain
\be
\omega^2 =  \ell^2 \tilde{U}_0 = \ell^2 \frac{f_0\, \alpha_0}{r^2_0}.
\label{omRgen}
\ee
Note that we could have arrived at the same result via the $\ell \gg 1$ limit of the standard WKB formula~\cite{SchutzWill}
which is still applicable for our wave equation with the assumed single-peak potential. In the GR limit of a Schwarzschild spacetime
we have $f = 1 -2M/r, ~ \alpha =1$, and our result reduces to the correct expression (see e.g.~\cite{Berti:2009kk}).

Writing the complex QNM frequency as
\be
\omega = \omega_R + i \omega_I,
\ee
we would normally expect $ \omega_R \gg |\omega_I|$ in the eikonal limit as in, e.g. Schwarzschild black holes. In that case
\be
\omega_R  \approx  \ell \, \tilde{U}_0^{1/2}.
\label{omReik1}
\ee
In principle, though, the leading-order result (\ref{omRgen}) should apply equally well for a QNM with the opposite arrangement
$|\omega_I| \gg \omega_R$ of real and imaginary parts.


\subsection{Eikonal approximation: subleading order}

Moving beyond the leading-order eikonal calculation, we expect  $\omega_I$ to be determined by the ${\cal O} (\epsilon^{-1})$ terms
of Eq.~(\ref{full_eik}) evaluated at $r=r_0$.  At this order we should also account for the subleading piece of $\omega_R$.  That is,
\be
\omega_R = \omega_R^{(0)} + \omega_R^{(1)} + {\cal O} (\ell^{-1}),
\quad  \omega_R^{(0)} = \ell \, \tilde{U}_0^{1/2},
\ee
where both $\omega_I,~ \omega_R^{(1)}$ are ${\cal O} (1)$ quantities.
Then,
\be
  \frac{2 i}{\epsilon} S_{,x} A_{,x} + \left ( \frac{i}{\epsilon} S_{,xx} +2 i \omega_R^{(0)} \omega_I  + 2\omega_R^{(0)} \omega_R^{(1)}
 - \ell \tilde{U} \right ) A = 0.
 \ee
 Setting $r=r_0$,
 \be
\frac{i}{\epsilon} S_{,xx} (r_0) +2 i \omega_R^{(0)} \omega_I + 2\omega_R^{(0)} \omega_R^{(1)} - \ell \, \tilde{U}_0 = 0.
 \label{sublead_eik}
\ee
To proceed, we must obtain $S_{,xx}$. By Taylor-expanding Eq.~(\ref{lead_eik}) around $r_0$ and making use of Eqs.~(\ref{dU0eq}) and (\ref{omReik1})
we find
\be
 \frac{(S_{,x})^2}{\epsilon^2}  = -\frac{\ell^2}{2}   \tilde{U}^{\pp}_0 (r-r_0)^2.
\ee
Taking the positive square root and then differentiating, we get
 \be
 \frac{S_{,xx}}{\epsilon}  =  \frac{\ell}{\sqrt{2}}   \frac{dr}{dx} | \tilde{U}^{\pp}_0 |^{1/2}.
 \label{SxxEq1}
 \ee
The imaginary part of (\ref{sublead_eik}) gives
\be
\omega_I =  -\frac{1}{2} \left (\frac{dr}{dx}\right )_0  \sqrt{\frac{| \tilde{U}^{\pp}_0 |}{2\tilde{U}_0}}.
\label{omIeik1}
\ee
As in the case of $\omega_R^{(0)}$, this $\omega_I$ result could have been derived by taking the $\ell \gg 1$
limit of a WKB formula.

Moving on, we can see that the real part of (\ref{sublead_eik}) furnishes the $\omega_R^{(1)}$ correction,
\be
\omega_R^{(1)} = \frac{1}{2}  \tilde{U}_0^{1/2}.
\ee
This means that we can write an expression for $\omega_R$ featuring the famous `$\ell +1/2$' Langer factor,
\be
\omega_R = \left (\ell + \frac{1}{2} \right )  \tilde{U}_0^{1/2} + {\cal O} (\ell^{-1}).
\label{omReik2}
\ee
Together with Eq.~(\ref{omIeik1}) this result completes our eikonal approximation analysis of the wave equation (\ref{gen_wave}).


\subsection{An eikonal QNM-photon geodesics connection?}
\label{sec:eik_geod}

As already discussed in some detail, the eikonal analysis of Schwarzschild (or Kerr) QNMs establishes a direct
connection between the fundamental QNM associated with the peak of the wave potential and the geodesic photon
ring~\cite{Ferrari:1984zz,Mashhoon:1985cya,Cardoso:2008bp, Dolan:2010wr, Berti:2009kk}. We can explore the validity
of this connection within the general setup of the previous section by assuming a Schwarzschild background spacetime
with $f=1-2M/r$.

The geodesic photon ring at $\rph=3M$ is the solution of $ 2f= r f^\p$ (see Appendix~\ref{sec:geodesics}
for a discussion of photon rings in a general spherical metric).
On the other hand, the peak $r_0$ of the eikonal wave potential solves the slightly different equation [see Eq.~(\ref{dU0eq})]
\be
r ( f\alpha)^\p = 2 f \alpha,
\label{dU0eqx}
\ee
and therefore  $r_0 \neq \rph$, \emph{unless} $\alpha=\mbox{const.}$ In this case
we inevitably face a breakdown of the eikonal QNM-photon ring correspondence [here we note that Ref.~\cite{Konoplya:2017wot}
considers higher-dimensional wave equations of the type (\ref{gen_wave}) and arrives at a similar conclusion].

As it turns out, however, the eikonal $\omega_R$ does `see' an \emph{effective} spherical metric with,
\be
-g^{\rm eff}_{tt} = \frac{1}{ g_{rr}^{\rm eff}} = f (r) \alpha (r).
\label{geffsingle}
\ee
Null geodesics in this metric obey the radial equation (see Appendix~\ref{sec:geodesics})
\be
(u^r)^2 = 1 - b^2 \tilde{U} (r)  = V_r (r,b),
\ee
where $b$ is the impact parameter. Circular photon orbits are singled out by the two conditions  $V_r =  V^\p_r = 0$; these
lead to the following equation for the `photon ring' radius [see Eq.~(\ref{rpheqsph2})],
\be
 r (g_{tt}^{\rm eff} )^\p = 2 g_{tt}^{\rm eff},
\label{rpheqsph2x}
\ee
which is identical to (\ref{dU0eqx}); therefore, the photon ring coincides with the peak $r_0$.
At the same time, the photon ring's `angular frequency' $\Omega_0$ is given by [this is Eq.~(\ref{Omph}) evaluated at
$\rph = r_0$]
\be
\Omega_0 = \frac{\sqrt{-g_{tt}^{\rm eff}(r_0)}}{r_0} = \frac{(f\alpha)^{1/2}_0}{r_0},
\label{Omphx}
\ee
and our earlier result (\ref{omReik1}) becomes $\omega_R \approx \ell \, \Omega_0$, i.e. the familiar relation between the
eikonal $\omega_R$ and the photon ring's angular frequency~\cite{Ferrari:1984zz,Mashhoon:1985cya,Cardoso:2008bp, Dolan:2010wr, Berti:2009kk}.

Encouraged by this result, the next objective is to examine whether the QNM's correspondence to an effective geodesic photon ring
extends to the imaginary part $\omega_I$ via a relation to the photon ring's  Lyapunov exponent.
The latter parameter is given by [see Ref.~\cite{Cardoso:2008bp} and Eq.~(\ref{Lyap1})]
\be
\gamma^2_0 = -\frac{1}{2} \, r^2_0 \, g_{tt}^{\rm eff} (r_0)  \left ( \frac{g_{tt}^{\rm eff}}{r^2} \right )^{\pp}_0.
\label{Lyap1x}
\ee
We can then see that (\ref{omIeik1}) becomes, after using the identification~\eqref{geffsingle},
\be
\omega_I = - \frac{1}{2} \left (\frac{dr/dx}{f \alpha}  \right )_0 |\gamma_0 |.
\ee
This reduces to the standard expression $\omega_I = - |\gamma_0 |/2 $ provided
\be
\frac{dr}{dx} = f \alpha = -g_{tt}^{\rm eff}.
\ee
Indeed, this is the same $x$ coordinate required for writing the scalar wave equation $ g_{\mu\nu}^{\rm eff} \nabla^\mu \nabla^\nu \Phi =0$
in the form (\ref{gen_wave}), see Appendix~\ref{sec:qnmgeod} for details.

We can thus conclude that the eikonal QNM of the generalised wave equation (\ref{gen_wave}), and in terms of the effective metric (\ref{geffsingle}),
retains the physical interpretation it enjoys in GR; i.e. it describes unstable null orbits in the photon ring of that metric.

In a slightly different scenario than the one considered here (see Appendix~\ref{sec:qnmgeod}), one can show that the eikonal QNM
of the usual scalar wave equation in a general spherical metric (and assuming that the present
section's model applies) is related to the properties of the metric's \emph{true} photon ring via the same leading-order relations
$\omega_R = \ell \, \Omega_{\rm ph},~\omega_I =  -|\gamma_{\rm ph} |/2$. In the light of this section's calculation this
result should not come as a total surprise given that the eikonal potential of the scalar wave equation (\ref{radscalar})
is $-\ell^2 \, g_{tt}/r^2$ in perfect analogy with this section's potential $U = -\ell^2 \, g_{tt}^{\rm eff}/r^2 $.


\section{Eikonal QNM of non-GR black holes}
\label{sec:nonGR}

\subsection{Coupled wave equations in a spherical black hole spacetime}
\label{sec:coupled}

The perturbation theory underpinning the calculation of QNMs of spherically symmetric black holes in modified theories of gravity
typically boils down to a coupled system of wave equations for the axial or polar components of the metric-tensor field ($\psi$) and the
additional field degrees of freedom; see for example~\cite{molina2010, Kobayashi:2012kh, Kobayashi:2014wsa, Blazquez-Salcedo:2016enn,Blazquez-Salcedo:2017txk,
Bertietal2015}. Our own discussion here is meant to be theory independent, but in order to keep this first analysis
simple, we assume a system comprising two perturbed fields $\psi$ and $\Theta$ (the scalar field, governed by a pair of coupled wave
equations,
\begin{subequations}
\begin{align}
&\frac{d^2 \psi}{dx^2} + \left [\, \omega^2 - V_\psi(r) \, \right ] \psi = \beta_\psi (r) \Theta,
\label{waveT}
 \\
& \frac{d^2 \Theta}{dx^2} + g(r)  \frac{d\Theta}{dx} + \left [\, \omega^2 - V_\Theta (r) \, \right ] \Theta = \beta_\Theta (r) \psi,
\label{waveS}
\end{align}
\end{subequations}
where $x$ is a common tortoise coordinate. The potential $V_\psi$ is assumed to be identical to the Schwarzschild's spacetime
Regge-Wheeler or Zerilli potential~\cite{ChandraBook} while the scalar potential $V_\Theta$ is allowed to deviate from GR,
\be
V_\psi = \{ V_{\rm RW}, V_{\rm Z} \},
\quad V_\Theta = f \left [ \frac{\ell(\ell+1)}{r^2}\alpha + \frac{2M}{r^3} \zeta  \right ],
\ee
with $f=1-2M/r$. The coupling functions $\beta_\psi(r)$,  $\beta_{\Theta} (r)$ and the potential functions
$\{g(r), \alpha(r), \zeta(r)\}$ are left undetermined, but we expect them to scale with negative powers of $r$ so that
asymptotic flatness is preserved\footnote{This requirement implies a massless scalar field.}.
We also assume that the $\ell (\ell +1)$ factor represents the entire $\ell$-dependence of the left-hand-side of (\ref{waveS}).
This `reduced' system is modeled after the perturbation equations for the axial tensor-scalar
perturbations of Schwarzschild black holes in Chern-Simons gravity~\cite{molina2010}, and it includes them as
a special case.

Our coupled system admits a QNM solution provided the appropriate boundary conditions are satisfied:
\be
\psi (x\to  \pm \infty) \sim e^{\pm i\omega x}, \quad \Theta (x\to  \pm \infty) \sim e^{\pm i\omega x}.
\ee
For this to be possible $\{ V_\psi$, $V_\Theta\}$ and $\{\beta_\psi, \beta_{\Theta}, g\}$ should vanish at $x = \pm \infty$.
The first set does indeed comply with this requirement provided $\alpha$ and $\zeta$ do not grow faster than $r^2$ and
$r^3$, respectively. The second set of parameters follows suit if these parameters scale with $f$ and negative powers of $r$.

With regard to approximation methods for a system like  (\ref{waveT})-(\ref{waveS}), and as far as we are aware,
no WKB formulae appear to exist in the literature. Fortunately, though, the eikonal/geometric optics approximation is flexible
enough to be applicable to a coupled system. To this end we assume the eikonal solution ansatz,
\be
\psi (x) = A(x) e^{iS(x)/\epsilon}, \quad \Theta (x) = B(x) e^{i H(x)/\epsilon},
\ee
where we have allowed for different phase functions for the two fields.

The best strategy for dealing with the coupled system at hand is to first combine the two equations and subsequently
take the eikonal limit of the resulting expression. As in the previous case of the single wave equation, it quickly becomes
evident that the two expansion parameters should balance, i.e.  $\epsilon \ell = {\cal O}(1)$, and that $\omega = {\cal O} (\ell)$.
Following this recipe, we obtain the following expression after having eliminated the $B$ amplitude between the two wave equations:
\begin{align}
& \omega^4 -\omega^2 \left [\,   \ell^2  \tilde{V}  (1+\alpha)  + \frac{1}{\epsilon^2} \left \{ (S_{,x})^2 + (H_{,x})^2 \right \} \, \right ] -\beta_{\psi\Theta}
\nn \\
&+  \ell^2  \tilde{V}   \left [  \alpha \left \{ \ell^2 \tilde{V}  + \frac{(S_{,x})^2}{\epsilon^2} \right \}  + \frac{(H_{,x})^2}{\epsilon^2}   \right ]
+ \frac{(S_{,x} H_{,x})^2}{\epsilon^4}
\nn \\
&   + \frac{\beta_\psi}{A} e^{i(H-S)/\epsilon} \left [ \left (g + \frac{2i}{\epsilon} H_{,x} \right ) B_{,x} + B_{,xx} \right ] + 2 \ell^3 \tilde{V}^2  \alpha
\nn \\
& - \frac{i}{\epsilon^3}  \left [\, \left ( g H_{,x}  +  H_{,xx} \right )  (S_{,x} )^2  + \left ( \frac{2 A_{,x}}{A} S_{,x}  + S_{,xx} \right ) (H_{,x})^2\, \right ]
\nn \\
&- \frac{i}{\epsilon} \ell^2 \tilde{V} \Big [\, \left ( \frac{2A_{,x}}{A} S_{,x} +  S_{,xx} \right ) \alpha + g H_{,x}  + H_{,xx}   \, \Big ]
\nn \\
&+  \frac{\ell \tilde{V}}{\epsilon^2} \left [ \alpha (S_{,x})^2 + (H_{,x})^2 \right ]
- \omega^2 \Big [\, \ell \tilde{V} (1+\alpha)  - \frac{i}{\epsilon} \Big ( g H_{,x}
\nn \\
& +  H_{,xx}  +  \frac{2 A_{,x}}{A} S_{,x}  + S_{,xx} \Big ) \,\Big ]  = 0  +  {\cal O} \left (\epsilon^{-2} \right),
\label{waveTSeik}
\end{align}
where we have defined $\beta_{\psi \Theta} \equiv \beta_\psi \beta_\Theta$ and $\tilde{V} (r) \equiv f(r)/r^2$. This expression includes all
terms up to ${\cal O}(\epsilon^{-3})$ order; all terms beyond that order have been omitted as they will not play any role in the subsequent analysis.
At the same time we have retained all $\beta$-coupling terms as a result of their unspecified eikonal order. From the above equation
it is clear that $\beta_{\psi\Theta} \leq {\cal O} (\ell^4)$, which is the highest allowed eikonal order consistent with the rest of the terms.
The presence of the exponential term would in principle be a cause for concern; assuming a $\beta_\psi \leq {\cal O} (\ell^2)$ pushes this
term into the group of ${\cal O} (\epsilon^{-3})$ subleading order terms where, as we shall see below, it makes no impact to the final results.
The same is true for the remaining amplitude-dependent terms.

Taking the eikonal limit,  $\epsilon \ll 1$ and $\ell \gg 1$, we have at leading order [i.e. the $ {\cal O} ( \epsilon^{-4})$ terms of (\ref{waveTSeik})]
\begin{align}
& \omega^4 -\omega^2 \left [\, \ell^2 \tilde{V}  (1+\alpha)  + \frac{1}{\epsilon^2} \left \{ (S_{,x})^2 + (H_{,x})^2 \right \} \, \right ]
\nn \\
&+ \ell^2  \tilde{V}   \left [  \alpha \left \{ \ell^2 \tilde{V}  + \frac{(S_{,x})^2}{\epsilon^2} \right \}  + \frac{(H_{,x})^2}{\epsilon^2}   \right ]
+ \frac{(S_{,x} H_{,x})^2}{\epsilon^4} = \beta_{\psi\Theta}.
\label{leadEq1}
\end{align}

The simplest situation is the one in which  $\beta_{\psi \Theta} \leq {\cal O} (\ell^3)$; in that
case the coupling term in (\ref{leadEq1}) is subdominant with respect to the other terms
and the system of the two equations effectively decouples. We discuss this case in more detail in the following section.


\subsection{The case of decoupled wave equations}

Here we focus on the scenario of  `weak $\ell$-coupling' where the term $\beta_{\psi\Theta} $ does not appear
in the leading-order eikonal equation (\ref{leadEq1}). This obviously includes the trivial case of \emph{no} coupling
$\beta_\psi = \beta_\Theta = 0$. Under these circumstances (\ref{leadEq1}) becomes the product of two factors,
\be
\frac{(S_{,x})^2}{\epsilon^2} = \omega^2 - \ell^2 \tilde{V}, \quad
\frac{(H_{,x})^2}{\epsilon^2} = \omega^2 -   \ell^2 \tilde{V}  \alpha.
\ee
Each of these equations is a special case of the general wave equation of Sec.~\ref{sec:wave_eik} and as a
consequence we can use the results obtained there.  For the $\psi$-field we recover the standard Schwarzschild result,
$\omega_R \approx \ell \, \tilde{V}_0^{1/2} =  \ell \, \Omega_{\rm ph}$ with $r_0 =\rph=3M$.
For the $\Theta$-field  we similarly obtain $\omega_R\approx \ell  \, ( \tilde{V} \alpha )_0^{1/2}$ with a different peak
$\tilde{r}_0 \neq 3M$ associated with the eikonal potential  $\ell^2 \tilde{V} \alpha$ where $H_{,x} =0$.

Clearly, due to the presence of the $\alpha(r)$ function the two eikonal frequencies do not coincide. This implies
that the two fields propagate independently, each one with its own QNM frequency.


\subsection{Leading-order analysis of the coupled system.}
\label{sec:fullcoupled}

The lesson from the preceding calculation is clear: the existence of a mixed scalar-tensor QNM wave
with a single frequency $\omega$ in the eikonal limit \emph{requires} the presence of a term
$\beta_{\psi \Theta} = {\cal O}(\ell^4)$ in the leading-order equation~(\ref{leadEq1}).

The next step is to assume that both phase functions are simultaneously minimised, $S_{,x}  = H_{,x} = 0$,
at the \emph{same} radius $r_{\rm m}$.  Provided $\beta_{\psi\Theta}$ is $\omega$-independent, the resulting expression
is a  biquadratic equation  in $\omega$:
\be
\omega^4 - \omega^2 \ell^2 \tilde{V}_{\rm m} ( 1+ \alpha_{\rm m})
+ \ell^4  \tilde{V}^2_{\rm m} \alpha_{\rm m}   -  (\beta_{\psi \Theta})_{\rm m} = 0,
\label{coupled5}
\ee
admitting the pair of roots
\begin{align}
\omega^2_{\pm} = \frac{\ell^2}{2}  \left [\, \tilde{V}_{\rm m} (1+\alpha_{\rm m})
\pm \sqrt{  \tilde{V}^2_{\rm m} (1-\alpha_{\rm m})^2 +  4 ( \tilde{\beta}_{\psi \Theta})_{\rm m}} \,  \right ],
\nn \\
\label{coupled6}
\end{align}
where $\tilde{\beta}_{\psi \Theta} = \beta_{\psi\Theta}/\ell^4$ is $\ell$-independent. Apart from its
self-consistent scaling, $\omega_{\pm} = {\cal O} (\ell)$, this result also has the correct GR limit
 ($\alpha \to 1,~ \beta_{\psi \Theta} \to 0 $) provided $r_{\rm m} \to 3M$.

As pointed out earlier, the eikonal $\omega_{\pm}$ may not necessarily coincide with the QNM with frequency
$\omega_R \gg |\omega_I |$. In the spirit of our theory-agnostic framework, we should not rule out modes with
the opposite arrangement $| \omega_I| \gg \omega_R $ or indeed ones with $| \omega_I | \sim \omega_R$ in the eikonal limit
[this last scenario requires at least one of the functions $\alpha, \beta_\psi, \beta_\Theta$ to be complex-valued and/or
$( \tilde{\beta}_{\psi \Theta})_{\rm m} < 0 $].

The radius $r_{\rm m}$ can be calculated by taking the derivative of (\ref{leadEq1}) and then evaluating it at
$(r,\omega)=(r_{\rm m}, \omega_\pm)$,
\be
\frac{\omega^2_\pm}{\ell^2} \left [  \tilde{V} ( 1+ \alpha)  \right ]^\p_{\rm m} - ( \alpha  \tilde{V}^2  )^\p_{\rm m}
+  (\tilde{\beta}_{\psi \Theta})^\p_{\rm m} = 0.
\label{coupled8}
\ee
This equation determines $r_{\rm m}$; in the GR limit  it reduces to $ \tilde{V}^\p =0$ and $ r_{\rm m} = \rph$.

Remarkably, $r_{\rm m} $ turns out to be a `peak' (in reality a local minimum or maximum)
of an \emph{effective} potential, albeit a frequency-dependent one. This can be easily verified by first defining
\be
V_{\rm eff}(r,\omega) \equiv \ell^2  \tilde{V} \left [   \omega^2 ( 1+ \alpha)  - \ell^2   \tilde{V} \alpha \right ] + \ell^4  \tilde{\beta}_{\psi \Theta},
\label{Veff}
\ee
and then noticing that (\ref{coupled8}) is equivalent to
\be
V_{\rm eff}^\p (r,\omega)=0, \quad \mbox{at}~ (r,\omega)=(r_{\rm m},\omega_\pm).
\label{dVeff}
\ee
With the effective potential $V_{\rm eff} (r, \omega)$ defined in this way, Eq. (\ref{coupled5}) becomes
\be
\omega^4_\pm = V_{\rm eff} (r_{\rm m}, \omega_\pm).
\label{omVeff}
\ee
Interestingly, one can define an alternative effective potential which is equivalent to (\ref{Veff}) in the sense
that both potentials satisfy the condition (\ref{dVeff}) and therefore have the same `peak' $r_{\rm m}$.
This second potential can be produced from (\ref{Veff}) after changing $\omega \to \omega_\pm(r)$, where
$\omega_\pm (r)$ is the function of Eq.~(\ref{coupled6}) with $r_{\rm m} \to r$.
For the same reason (\ref{omVeff}) becomes $\omega^4_\pm (r) = V_{\rm eff} (r,\omega_\pm (r))$. Then,
\be
\frac{d}{dr}  V_{\rm eff} (r,\omega_\pm (r)) = V_{\rm eff}^\p + V_{\rm eff,\omega_\pm}\, \omega^\prime_\pm,
\ee
and at the same time,
\be
\\
4 \omega^3_\pm \omega^\prime_\pm = \frac{d}{dr}  V_{\rm eff} (r,\omega_\pm (r)).
\ee
The combination of these two equations implies $V_{\rm eff}^\p =0 \Leftrightarrow \omega^\p_\pm =0$,
thus ensuring the equivalence of the two potentials.


\subsection{The coupled system at subleading eikonal order}
\label{sec:coupled_sub}

Going beyond the leading eikonal order, we assume a `canonical' QNM with $\omega_R \gg |\omega_I | $
and
\be
\omega = \omega_R^{(0)} + \omega_R^{(1)} + i \omega_I +  {\cal O} (\ell^{-1}), \quad \omega_R^{(0)} = \omega_\pm.
\ee
We also expand the coupling parameter as
\be
\beta_{\psi\Theta} =  \beta_{\psi\Theta}^{(0)} +  \beta_{\psi\Theta}^{(1)}+ {\cal O}(\ell^2), \quad
 \beta_{\psi\Theta}^{(0)} = \ell^4 \tilde{\beta}_{\psi\Theta}.
\ee
The subleading order eikonal equation consists of the ${\cal O} (\epsilon^{-3})$ terms of the general expression (\ref{waveTSeik}).
After setting $r=r_{\rm m}$, all terms with $S_{,x}$ and $H_{,x}$ factors vanish (including terms with the `frictional' parameter $g$
and exponentials) and we obtain
\begin{align}
&   2 \omega_\pm \omega_R^{(1)} \left [\,  2 \omega^2_\pm -  \ell^2 \tilde{V}_{\rm m} (1+ \alpha_{\rm m}) \, \right ]
-\ell \tilde{V}_{\rm m}\Big [\, \omega_\pm^2   (1+ \alpha_{\rm m} )
\nn \\
&- 2 \ell^2 \tilde{V}_{\rm m}  \alpha_{\rm m}  \Big ] - (\beta_{\psi\Theta}^{(1)})_{\rm m}
+ i \Big [ \,  (H_{,xx})_{\rm m} \left ( \omega^2_\pm -\ell^2 \tilde{V}_{\rm m} \right )
 \nn \\
&  + (S_{,xx})_{\rm m} \left ( \omega^2_\pm -  \ell^2 \tilde{V}_{\rm m}  \alpha_{\rm m} \right )
+ 2 \omega_\pm \omega_I \Big \{\,  2 \omega^2_\pm
\nn \\
& -  \ell^2 \tilde{V}_{\rm m} (1+ \alpha_{\rm m}) \, \Big \}  \, \Big ]  = 0.
\end{align}
We decompose this into real and imaginary parts, assuming for simplicity an entirely real $\beta_{\psi\Theta}$,
\begin{subequations}
\begin{align}
&   2 \omega_\pm \omega_R^{(1)} \left [\,  2 \omega^2_\pm -  \ell^2 \tilde{V}_{\rm m} (1+ \alpha_{\rm m}) \, \right ]
-\ell \tilde{V}_{\rm m}\Big [\, \omega_\pm^2   (1+ \alpha_{\rm m} )
\nn \\
&- 2 \ell^2 \tilde{V}_{\rm m}  \alpha_{\rm m}  \Big ] - (\beta_{\psi\Theta}^{(1)})_{\rm m} = 0,
\label{subleadEqR}
\\
\nn \\
& 2 \omega_\pm \omega_I \left [\,  2 \omega^2_\pm  -  \ell^2 \tilde{V}_{\rm m} (1+ \alpha_{\rm m}) \, \right ]
+ (H_{,xx})_{\rm m} \left ( \omega^2_\pm -\ell^2 \tilde{V}_{\rm m} \right )
\nn \\
&+ (S_{,xx})_{\rm m} \left ( \omega^2_\pm -  \ell^2 \tilde{V}_{\rm m}  \alpha_{\rm m} \right ) = 0.
\label{subleadEqIm}
\end{align}
\end{subequations}
This is a decoupled system of equations for the subleading frequency corrections $\{\omega_R^{(1)},\omega_I \}$
with roots
\begin{align}
& \omega_I =  \frac{(H_{,xx})_{\rm m} \left ( \omega^2_\pm  -\ell^2 \tilde{V}_{\rm m} \right )
+ (S_{,xx})_{\rm m} \left [ \omega^2_\pm - \ell^2 \tilde{V}_{\rm m}  \alpha_m  \right ]}
{2\omega_\pm \left [\,  \ell^2 \tilde{V}_{\rm m} (1+ \alpha_{\rm m} ) -2 \omega^2_\pm \,\right ]},
\label{omIroot1}
\\
\nn \\
& \omega_R^{(1)} = \frac{ \ell \tilde{V}_{\rm m} \left [\, \omega^2_\pm (1+\alpha_{\rm m})
- 2\ell^2 \tilde{V}_{\rm m}^2 \alpha_{\rm m}  \, \right ] + (\beta^{(1)}_{\psi\Theta})_{\rm m} }{2\omega_\pm \left [\, 2 \omega^2_\pm
-  \ell^2\tilde{V}_{\rm m} (1+ \alpha_{\rm m} ) \,\right ]}.
\label{omR1root1}
\end{align}
In order to calculate the second derivatives $\{ S_{,xx}, H_{,xx}\}$ we rewrite (\ref{leadEq1}) in the equivalent form
\begin{align}
& \omega^4 -V_{\rm eff} (r,\omega)  - \frac{(S_{,x})^2}{\epsilon^2}  \left ( \omega^2 - \ell^2 \tilde{V} \alpha \right )
 \nn \\
& -\frac{(H_{,x})^2}{\epsilon^2} \left (  \omega^2 - \ell^2 \tilde{V} \right ) + \frac{(S_{,x} H_{,x})^2}{\epsilon^4} = 0,
\end{align}
and then expand around $r=r_{\rm m}$. With the help of
\begin{align}
S_{,x} (r) &\approx \left ( \frac{dx}{dr} \right )_{\rm m} (S_{,xx})_{\rm m} (r-r_{\rm m}),
\\
H_{,x} (r) &\approx  \left ( \frac{dx}{dr} \right )_{\rm m} (H_{,xx})_{\rm m} (r-r_{\rm m}),
\\
V_{\rm eff} (r,\omega_\pm) & \approx \omega^4_\pm  + \frac{1}{2} V_{\rm eff}^{\pp}  (r_{\rm m}, \omega_\pm) (r-r_{\rm m} )^2,
\end{align}
we obtain at leading order in $r-r_{\rm m}$,
\begin{align}
& \frac{(S_{,xx})^2_{\rm m}}{\epsilon^2} \left ( \omega^2_\pm -\ell^2 \tilde{V}_{\rm m}  \alpha_{\rm m} \right )
 + \frac{(H_{,xx})^2_{\rm m}}{\epsilon^2} \left ( \omega^2_\pm - \ell^2 \tilde{V}_{\rm m} \right )
\nn \\
& =  - \frac{1}{2}   \left ( \frac{dr}{dx} \right )_{\rm m}^2 V_{\rm eff}^{\pp}  (r_{\rm m}, \omega_\pm).
\label{coupled9}
\end{align}
As it stands, this expression cannot be manipulated further unless we assume a relation between the
derivatives of the phase functions. The simplest choice is to set $ (S_{,xx})^2_{\rm m} = (H_{,xx})^2_{\rm m}$;
this would clearly be the case if the phase functions were equal (up to a constant)\footnote{The equality between the phase functions
would imply a common wave propagation speed for the two fields.}, i.e. $S_{,x} =H_{,x}$. With this assumption (\ref{coupled9}) becomes
\begin{align}
& \left [ 2 \omega^2_\pm  -\ell^2 \tilde{V}_{\rm m} (1+  \alpha_{\rm m} )  \right ] \frac{(S_{,xx})^2_{\rm m} }{\epsilon^2}
=  \pm \ell^2 \Big [  \tilde{V}_{\rm m}^2 (1-  \alpha_{\rm m} )^2
\nn \\
& +  4 ( \tilde{\beta}_{\psi \Theta} )_{\rm m}   \Big ]^{1/2} \frac{(S_{,xx})^2_{\rm m} }{\epsilon^2}
= -\frac{1}{2}  \left ( \frac{dr}{dx} \right )_{\rm m}^2 V_{\rm eff}^{\pp} (r_{\rm m}, \omega_\pm ).
\label{coupled10}
\end{align}
Assuming a potential peak, $V_{\rm eff}^{\pp} (r_{\rm m}, \omega_\pm ) < 0 $, it is always possible to choose the sign
of the square root so that $(S_{,xx})_{\rm m} >0$. We then have
\begin{align}
\frac{(S_{,xx})_{\rm m}}{\epsilon} &=   \frac{1}{\sqrt{2} \ell} \left (\frac{dr}{dx} \right )_{\rm m}
 | V_{\rm eff}^{\pp} (r_{\rm m}, \omega_\pm ) | ^{1/2}
 \nn \\
&\quad \times \left [ \tilde{V}_{\rm m}^2 (1-\alpha_{\rm m})^2 + 4 ( \tilde{\beta}_{\psi\Theta} )_{\rm m}   \right ]^{-1/4},
\end{align}
and solution (\ref{omIroot1}) for $\omega_I$ becomes
\begin{align}
\omega_I & = - \frac{(S_{,xx})_{\rm m}}{2 \omega_\pm}
\nn \\
&= -\frac{(dr/dx)_{\rm m}}{2 \sqrt{2} \omega_\pm \ell}  \frac{  | V_{\rm eff}^{\pp} (r_{\rm m}, \omega_\pm ) |^{1/2}}
{ \left [\,  \tilde{V}_{\rm m}^2 (1-\alpha_{\rm m})^2 + 4 ( \tilde{\beta}_{\psi \Theta} )_{\rm m}  \, \right ]^{1/4} }.
\label{omIroot2}
\end{align}

\subsection{Summary of eikonal formulae}
\label{sec:summary}

Here we recap the eikonal results of this section for the complex QNM frequency $\omega = \omega_R + i \omega_I$
of the coupled system (\ref{waveT}) and (\ref{waveS}).
We have
\be
\omega = \omega_R^{(0)} + \omega_R^{(1)} + i \omega_I +  {\cal O} (\ell^{-1}),
\ee
with
\begin{align}
 \omega_R^{(0)}  &= \omega_{\pm} = \frac{\ell}{\sqrt{2}}  \Big [\, \tilde{V}_{\rm m} (1+\alpha_{\rm m})
 \nn \\
&\quad\quad\quad\quad \pm \sqrt{  \tilde{V}^2_{\rm m} (1-\alpha_{\rm m})^2 +  4 ( \tilde{\beta}_{\psi \Theta})_{\rm m}} \,  \Big ]^{1/2},
\label{omR0_final}
\\
\nn \\
\omega_R^{(1)} &= \frac{ \ell \tilde{V}_{\rm m} \left [\, \omega^2_\pm (1+\alpha_{\rm m})
- 2\ell^2 \tilde{V}_{\rm m}^2 \alpha_{\rm m}  \, \right ] + (\beta^{(1)}_{\psi\Theta})_{\rm m} }{2\omega_\pm \left [\, 2 \omega^2_\pm
-  \ell^2\tilde{V}_{\rm m} (1+ \alpha_{\rm m} ) \,\right ]},
\label{omR1_final2}
\end{align}
for the real part and
\be
\omega_I = -\frac{(dr/dx)_{\rm m}}{2 \sqrt{2} \omega_\pm \ell}  \frac{  | V_{\rm eff}^{\pp} (r_{\rm m}, \omega_\pm ) |^{1/2} }
{ \left [\,  \tilde{V}_{\rm m}^2 (1-\alpha_{\rm m})^2 + 4 ( \tilde{\beta}_{\psi\Theta} )_{\rm m}  \, \right ]^{1/4} },
\label{omI_final}
\ee
for the imaginary part. For the coupling functions we have assumed $\beta_\psi \leq {\cal O} (\ell^2)$
and that $\beta_{\psi\Theta} = \beta_\psi \beta_\Theta $ is real-valued, $\omega$-independent and with an $\ell \gg 1$
expansion $ \beta_{\psi\Theta} = \ell^4   \tilde{\beta}_{\psi\Theta} + \beta_{\psi\Theta}^{(1)} + {\cal O}(\ell^2) $.
The potentials appearing in these expressions are given by
\begin{align}
    V_{\rm eff} (r,\omega) & =  \ell^2  \tilde{V} \left[   \omega^2 ( 1+ \alpha)  - \ell^2   \tilde{V} \alpha \right] + \ell^4  \tilde{\beta}_{\psi \Theta},
\label{Veff_final}
\\
\tilde{V} (r) &= \frac{f}{r^2},
\end{align}
and the potential peak radius $r_{\rm m}$ solves $V_{\rm eff}^\p (r_{\rm m},\omega_{\pm})=0$.
It is straightforward to verify that in the GR limit,
\be
 V_{\rm eff}^{\pp} (r_{\rm m}, \omega_\pm ) \to 2 \ell^2 \left [\, \tilde{V}_{\rm m}^{\pp} (\omega^2_\pm -\ell^2 \tilde{V}_{\rm m})
 -\ell^2 (\tilde{V}^\p_{\rm m})^2 \, \right ],
\ee
and $ \tilde{V}^\p_{\rm m} \to 0$. Then, the above results reduce to the familiar expressions
\begin{align}
& \omega_R^{(0)} \to \ell \tilde{V}_{\rm m}^{1/2}, \quad
\omega_R^{(1)}  \to \frac{1}{2} \tilde{V}_{\rm m}^{1/2},
\\
&\omega_I  \to  -\frac{1}{2} \left (\frac{dr}{dx} \right )_{\rm m} \sqrt{ \frac{|\tilde{V}^{\pp}_{\rm m}|}{ 2\tilde{V}_{\rm m}} },
\end{align}
with $r_{\rm m} \to 3 M$.


\subsection{A `photon ring' for the coupled system}
\label{coupled_geod}

The emergence of $V_{\rm eff}$ in the eikonal calculation of the preceding sections prompts us to explore the possibility of
a `geodesic' connection to an effective photon ring. The form of $V_{\rm eff}$  and the correspondence
$\ell/\omega \to b$ suggests an effective radial geodesic potential,
\be
V_r^{\rm eff} (r,b) = 1 -  b^2  \left [\, \tilde{V}(1 + \alpha) - b^2 \tilde{V}^2 \alpha -  b^2 \tilde{\beta}_{\psi\Theta} \,\right ].
\label{Vreff}
\ee
The clear dissimilarity to the radial potential (\ref{pheom}) of null geodesics in a spherical metric suggests that we should
not be too optimistic about attaching a geodesic analogy to the eikonal QNM of the coupled system.

A `photon ring' in the potential (\ref{Vreff}) is defined by
\be
V_r^{\rm eff} = 0, \quad (V_{r}^{\rm eff} )^\p = 0.
\ee
The first condition leads backs to $b_\pm^2 = \ell^2/\omega_{\rm \pm}^2 $ and then the second one can be identified with Eq.~(\ref{coupled8})
for $r_{\rm m}$. In other words, $r_{\rm m }$ is the photon ring radius of the `geodesic' potential $V^{\rm eff}_r$.

We can rewrite the impact parameter result as,
\be
\omega_\pm  = \ell \, b_\pm^{-1} \equiv \ell\, \Omega_\pm,
\ee
in accordance with the definition (\ref{Omph}) of angular frequency in spherical symmetry.
Thus defined, $\Omega_\pm$ represents the effective photon ring's angular frequency.

This is as far as the geodesic analogy can be pushed.  Although a Lyapunov exponent can be defined for the potential
(\ref{Vreff}), we find that it is not related to the $\omega_I$ given by (\ref{omI_final}) in the same way as in the
single wave equation case.
With our eikonal analysis of the coupled system brought to an end we are now ready to study a specific example of
a modified theory of gravity.


\section{Application: Schwarzschild black holes in Chern-Simons gravity}
\label{sec:CS}

Chern-Simons gravity (dCS), in its dynamical version, represents an extension of GR achieved by adding a parity-violating term to
the standard Einstein-Hilbert action; see \cite{jackiw2003,alexander2009, yunes2009} for further details. As a result of this modification
the gravitational field in this theory acquires a dynamical scalar field degree of freedom in addition to the standard tensorial one.
As far as black holes are concerned, the theory admits the Schwarzschild metric as a spherically symmetric vacuum solution with a
vanishing background scalar field~\cite{Yunes2008,Cardoso:2009pk}. The polar perturbations of these black holes are described by the
familiar general relativistic Zerilli equation~\cite{ChandraBook} while the axial sector ($\psi$) of the perturbations couples to the perturbed
scalar field ($\Theta$). This coupling signals the breakdown of isospectrality between the polar and axial QNM sectors; the former modes
remain the same as in GR while the latter are governed by a system of coupled wave equations of the form (\ref{waveT}) and (\ref{waveS})
with~\cite{Cardoso:2009pk, molina2010}
\begin{subequations}
\begin{align}
& V_\psi = V_{\rm RW} = f \left [ \frac{\ell (\ell+1)}{r^2} -\frac{6M}{r^3}\right ], \quad g = 0,
\\
&  \frac{dr}{dx} = f = 1 - \frac{2M}{r}, \quad \alpha =1 + \frac{576 \pi M^2}{\beta r^6}, \quad \zeta =1,
\\
& \beta_\psi = 96  \pi M \frac{f }{r^5}, \quad \beta_\Theta =  \frac{6 M}{ \beta} \frac{(\ell+2)!}{(\ell-2)!} \frac{f}{r^5},
 \end{align}
\end{subequations}
where $\beta$ is the theory's coupling constant (GR is formally recovered in the limit $ M^4 \beta \to \infty$).
We can observe that the dependence of these functions on $\{r,\ell, f\}$ is consistent with the constraints discussed
in Secs.~\ref{sec:coupled} and \ref{sec:fullcoupled} in relation to the necessary QNM boundary conditions and
the $\ell$-scaling of $\beta_{\psi\Theta}$.

The aim of this section is to compare the numerically computed axial tensor-scalar QNMs of dCS black holes
reported in Ref.~\cite{molina2010} to the leading-order eikonal formulae (\ref{omR0_final}) and (\ref{omI_final}).
We limit our discussion to the more physically relevant case\footnote{A $\beta < 0$ leads to a negative kinetic energy term
in the action and as a result the theory is infested with ghostlike instabilities.} $\beta > 0$ \cite{molina2010}.

The eikonal limit of the coupling functions,
\be
\beta_{\psi\Theta} = \frac{576 \pi M^2}{\beta} \frac{f^2}{r^{10}} \left [ \ell^4 + 2 \ell^3 -\ell^2 + {\cal O}(\ell) \right ],
\ee
allows for easy identification of $\tilde{\beta}_{\psi \Theta}$ and $\beta_{\psi\Theta}^{(1)}$.
For the real part of the eikonal QNM, we find
\begin{align}
\omega^2_\pm =  \frac{\ell^2 f_{\rm m}}{r^2_{\rm m}} \left[  1 +   \frac{288 \pi M^2}{\beta r_{\rm m}^6}
\left (\, 1 \pm  \sqrt{  1+ \frac{\beta r^6_{\rm m} }{ 144 \pi M^2} } \, \right ) \right].
\nonumber \\
\label{omCS}
\end{align}
The expression for $\omega_I$ is somewhat cumbersome and, as a consequence, is not shown here.
Equation~(\ref{coupled8}) for the potential peak radius $r_{\rm m}$ takes the form
\begin{align}
&  \Big [ 288 \pi M^2 \left ( 4 r_{\rm m}  -9 M  \right ) +  \beta r^6_{\rm m}   (r_{\rm m}  - 3 M )  \Big ]
\nn \\
& \times \Big [ 288 \pi M^2   + \beta r^6_{\rm m}  \pm 24M \sqrt{\pi(144 \pi M^2  + \beta r^6_{\rm m}  )}  \Big ]
 \nn \\
& - r^{12}_{\rm m}   (r_{\rm m} -3M) \beta^2 = 0.
\label{rmCS}
\end{align}
The fact that the right-hand side of (\ref{omCS}) is real  implies $\omega_R \gg | \omega_I|$ or
$| \omega_I | \gg \omega_R$ for the eikonal modes. Considering the strong `anti-GR' coupling limit $ M^4 \beta \ll 1$,
we find the asymptotic expressions,
\be
\omega_{+} \approx \frac{ 2048\ell }{6561} \sqrt{\frac{\pi}{\beta}},
\quad r_{\rm m} \approx \frac{9}{4} M \left ( 1 + \frac{19683}{524288} \frac{\beta}{\pi} \right ),
\ee
for the `+ mode' and
\be
\omega_{-} \approx \frac{\ell f_{\rm m}^{1/2}}{r_{\rm m}} \left [ 1 -12 \left (\frac{\pi M^2}{\beta r_{\rm m}^6} \right )^{1/2} \right ],
\quad \frac{\beta r_{\rm m}^6}{M^2} \gg 1,
\ee
for the `- mode'.  The divergence of $\omega_+$ as $\beta \to 0$ suggests that this mode may not be physically relevant in the strong
coupling regime.

Table~\ref{tab:CSqnms} collects numerical QNM data from Ref.~\cite{molina2010} and our eikonal frequencies
for the same values of $\beta$ and $\ell$. In all cases Eq.~(\ref{rmCS}) leads to a unique real solution with $r_{\rm m} > 2M$.
Starting from the intermediate coupling regime, $M^4 \beta \sim 1$, we find good agreement between the two sets of results
with an overall precision ($\sim 10\%-20\%$) which is typical of the eikonal approximation in the low-$\ell$ regime.
We can observe that among the two eikonal modes, $\omega_{-}$ is the least damped and the one that lies closer to a numerical QNM.
The precision of the $\omega_I$ results can be seen to be far better than that of $\omega_R$; in the examples
shown here the agreement can extend to two or three significant digits!

Moving towards the GR limit,  $M^4 \beta \gg 1$, the two eikonal modes approach each other, while the numerical modes converge to the usual
QNM frequencies of decoupled gravitational and scalar perturbations in the Schwarzschild spacetime (while, at the same time, $r_{\rm m} \to 3M$).
Interestingly, the eikonal $\omega_R$ ($\omega_I$) is seen to converge towards the corresponding component of the gravitational (scalar) QNM frequency.

\begin{figure*}[htb!]
\includegraphics[width=0.49\textwidth]{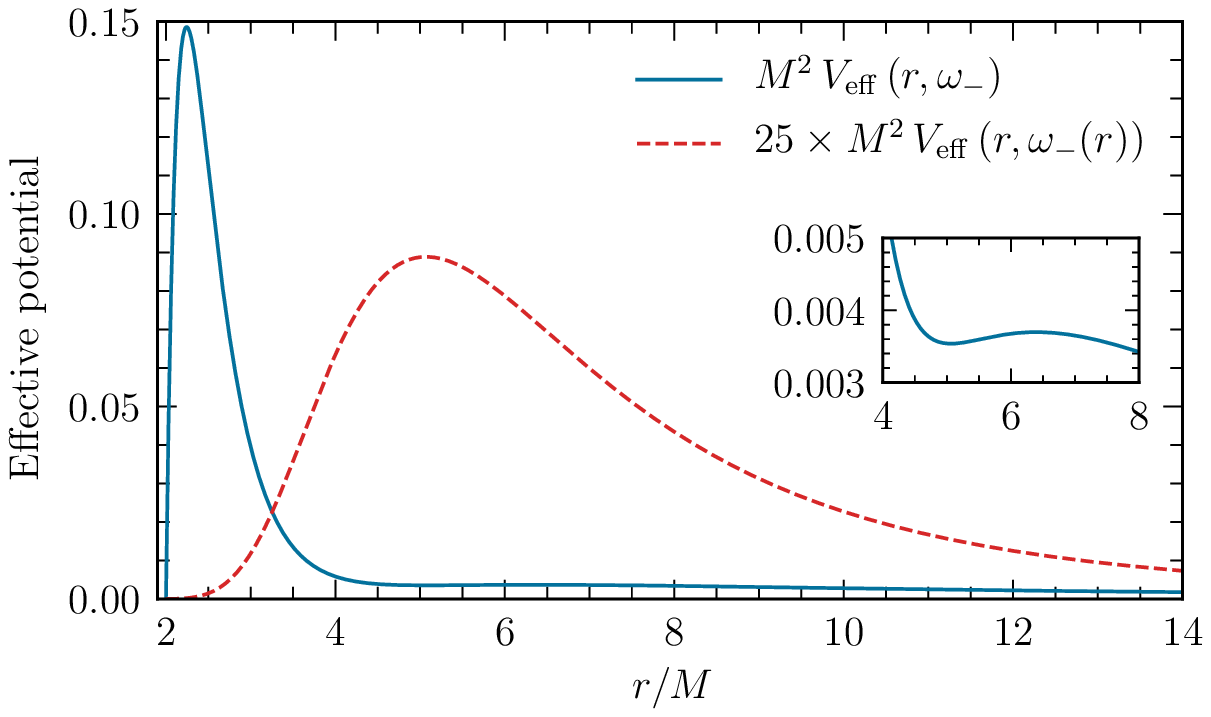}
\includegraphics[width=0.49\textwidth]{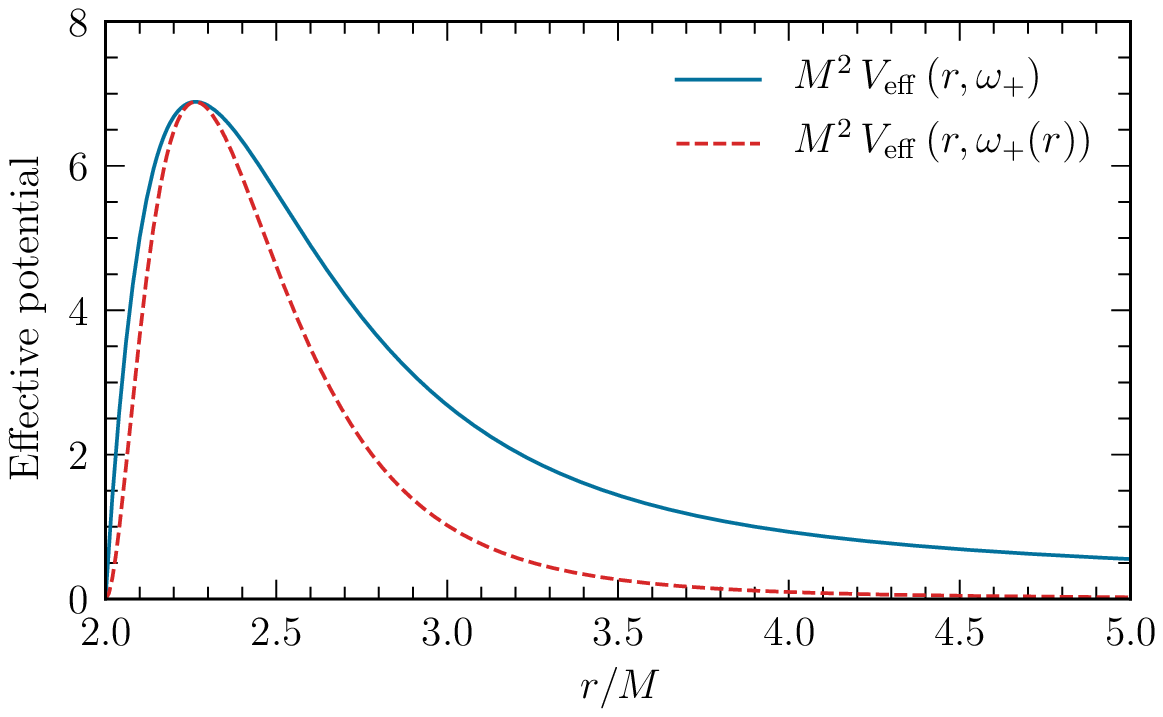}
\caption{\emph{The effective potentials for axial tensor-scalar QNMs of Schwarzschild black holes in dCS gravity}.
We show the $M^4\beta=0.5,~\ell=2$ potentials $V_{\rm eff} (r, \omega_\pm)$ and $V_{\rm eff}(r, \omega_\pm (r))$ defined in Sec.~\ref{sec:fullcoupled}.
The left (right) panel corresponds to the $-$ ($+$) mode solution. The corresponding $r_{\rm m}$ solutions (cf. Table~\ref{tab:CSqnms}) mark the common
extrema of the two potentials. In the right panel, these extrema are both maxima, while in the left panel (see inset), it is a maximum for $V_{\rm eff} (r, \omega_{-}(r))$
and a minimum for $V_{\rm eff} (r, \omega_{-})$.
}
\label{fig:VeffCS}
\end{figure*}

\begin{table}[t]
\begin{center}
\begin{tabular}{c c c c c c c}
\hline\hline
$ M^4 \beta$      & $\ell$  & $+/-$  & $ r_{\rm m} /M  $  & $ M \omega_{\rm eik}  $  & $ M \omega_{\rm num} $ & $\Delta \omega$ [\%]  \\
\hline
0.005             &  2     & $-$   &  9.96       &  0.133-0.0642$i$    & 0.186-0.0606$i$   & 39.8+5.6$i$ \\
                  &  2     & $+$   &  2.25       &  15.7-0.148$i$     &                    &  \\
                  & 10     & $-$   &  9.96       &  0.665-0.0642$i$    & 0.696-0.0636$i$   & 4.7+0.9$i$\\
                  & 10     & $+$   &  2.25       &  78.3-0.14$i$      &                    & \\
\hline
0.04              & 2      & $-$   &  7.24       &  0.178-0.0793$i$    & 0.220-0.0760$i$  & 23.6+4.1$i$  \\
                  & 2      & $+$   &  2.25       &  5.55-0.148$i$     &                     \\
                  & 7      & $-$   &  7.24       &  0.624-0.0793$i$    & 0.662-0.0687$i$ & 6.1+13.4$i$   \\
                  & 7      & $+$   &  2.25       &  19.4-0.148$i$     &                     \\
\hline
0.5               & 2      & $-$   &  5.07       &  0.244-0.0936$i$    & 0.276-0.0936$i$  & 13.1+0$i$ \\
                  &        & $+$   &  2.26       &  1.620-0.144$i$     & 1.97-0.144$i$   & 21.6+0$i$ \\
\hline
1                 & 2      & $-$   &  4.65       &  0.263-0.0959$i$    & 0.292-0.0971$i$  & 11.0+1.3$i$  \\
                  &        & $+$   &  2.28       &  1.183-0.141$i$     & 1.43-0.142$i$   & 20.9+0.7$i$   \\
\hline
100               & 2      & $-$   &  3.23       &  0.359-0.0968$i$   & 0.367-0.092$i$    & 2.2+4.9$i$ \\
                  &        & $+$   &  2.77       &  0.421-0.0976$i$    & 0.501-0.0954$i$  & 19.0+2.3$i$   \\
\hline
$10^4$            & 2      & $-$   &  3.02       &  0.382-0.0962$i$    & 0.374-0.0889$i$  & 2.1+7.6$i$  \\
                  &        & $+$   &  2.98       &  0.388-0.0962$i$    & 0.484-0.0968$i$  & 24.7+0.6$i$  \\
\hline\hline
\end{tabular}
\end{center}
\caption{\emph{Axial tensor-scalar QNMs of a Schwarzschild black hole in dCS gravity}.  We show
eikonal results ($M \omega_{\rm eik}$) corresponding to the modes $\omega_{\pm}$
computed with the help of Eqs.~(\ref{omI_final}), (\ref{omCS}) and (\ref{rmCS}) against
numerical data ($M \omega_{\rm num}$)~\cite{molina2010,molina_private} for a variety of coupling strengths $M^4 \beta$ and multipoles $\ell$.
The relative error $\Delta \omega = | [ (\omega_{\rm eik})_{R,I} -(\omega_{\rm num})_{R,I}]/(\omega_{\rm eik})_{R,I}| $ is shown as a complex
number in the last column. We also tabulate the radial location $r_{\rm m}$ of the `peak' of the effective potential $V_{\rm eff}$, as obtained
from (\ref{rmCS}).}
\label{tab:CSqnms}
\end{table}
In the opposite limit of strong coupling, $M^4 \beta \ll 1$, Ref.~\cite{molina2010} finds that the black hole's ringdown is dominated
by nonoscillatory QNMs. At first glance, this seems to be at odds with the eikonal formulae's prediction of oscillatory modes
(no matter which of $\omega_R, \omega_I$ is the dominant component). As it turns out, however, these nonoscillatory modes do not
represent the black hole's fundamental mode. Indeed, nonoscillatory modes also appear in the late part of the intermediate regime
($M^4 \beta \sim 1$) time evolutions, immediately after the ringdown of the fundamental mode, see Fig. 2 in~\cite{molina2010}.
As $\beta$ is reduced,  the nonoscillatory mode emerges at an earlier stage thus dominating most of the time domain signal.
A recent unpublished (and preliminary) QNM calculation~\cite{molina_private} reveals the presence of oscillatory modes
in the time domain signal of perturbed, strong coupling regime black holes; this are the data displayed in the top two rows of
Table~\ref{tab:CSqnms} for $M^4 \beta = \{0.005, 0.04\}$.  The $\ell=2$ eikonal frequency exhibits a deteriorated precision
($\sim 20\% - 40 \%$) with respect to the precision in the $M^4 \beta \gtrsim 1$ regime. On the other hand, the accuracy of the
eikonal damping rate is much better ($\sim 5 \%$). As expected, the accuracy of the eikonal results gets better for the
higher $\ell$ multipoles. It is also likely that the overall precision of the eikonal approximation in the $M^4 \beta \ll 1$ regime
may get better when more accurate numerical QNM results become available.

Besides the frequencies themselves, it is interesting to study the form of the effective potential $V_{\rm eff}$ as defined in two
equivalent ways at the end of Sec.~\ref{sec:fullcoupled}. Figure~\ref{fig:VeffCS} displays a typical example of this potential for the
$M^4 \beta=0.5$, $\ell=2$ QNMs appearing in Table~\ref{tab:CSqnms}. The shape of the potential depends on which definition we adopt, namely,
$V_{\rm eff} (r,\omega_\pm)$ or $V_{\rm eff} (r,\omega_\pm (r))$. In this particular example, three out of the four potentials are found to be
black hole-like, with a single hump located at the same $r_{\rm m}$ for a given mode. In contrast, the remaining fourth potential has a local
minimum at $r_{\rm m}$ and two maxima at different radii (see inset in Fig.~\ref{fig:VeffCS}). In all cases $V_{\rm eff} \to 0$ as $r\to \{2M, \infty\}$
in agreement with the QNM boundary conditions. Moreover, by taking the $M^4 \beta \to \infty$ limit in (\ref{Veff_final}),
the effective potential converges to $\omega^2 \ell^2 \tilde{V}$.


\section{Concluding remarks}
\label{sec:conclusions}

The main results of this paper, Eqs.~(\ref{omR0_final})--(\ref{omI_final}), represent the leading-order
eikonal QNM complex frequency for black hole perturbations described by the general coupled system of
wave equations (\ref{waveT}) and (\ref{waveS}). Furthermore, this eikonal mode (which is an approximation
to the black hole's fundamental mode) can be associated with the peak of a single effective potential, Eq.~(\ref{Veff_final}),
in much the same way as it happens for the eikonal modes of Schwarzschild or Kerr black holes in GR~\cite{Berti:2009kk}.
As a performance benchmark of our results, we have computed eikonal modes of Schwarzschild black holes in dCS gravity and found
them to be in good agreement with numerically computed data~\cite{molina2010, molina_private}.

The strength of the eikonal method lies in the analytic form of its results and the present study is no exception to the rule.
Although the QNM spectrum of spherical black holes in a given modified theory of gravity could, in principle, be obtained by means
of direct numerical integration, our eikonal formulae have the key merit of describing the fundamental mode of black holes in a
parametrised `post-GR' form and explicitly displaying the mode's dependence on the coupling parameters that generically
appear in modified theories of gravity.
In addition, our framework is sufficiently general to cover background black hole solutions which are either
given analytically or numerically.

Our exploration of the connection between the eikonal QNMs and photon geodesics has been partially successful.
We have been able to identify an effective metric and photon ring that can be mapped on the QNM of a general class
of single-field wave equations. The more complicated (and physically relevant) coupled system of equations has only allowed us to
associate an effective potential and its peak to the eikonal QNM. However, the connection of this potential to the true photon
geodesics of a given theory is an open question.

Several gravity theories lead to more complex perturbation systems than the one considered here, featuring
more than two wave equations/perturbed fields (including massive scalar fields) and/or higher-order derivative terms (see for
example~\cite{Blazquez-Salcedo:2016enn, Blazquez-Salcedo:2017txk, Brito:2018hjh, Tattersall_2018a, Tattersall_etal2018}).
In principle, the eikonal scheme of this paper should be adaptable to these more general scenarios at the cost of an increased
algebraic complexity. An equally important extension of this work -- and one that could lift our capability of testing the Kerr hypothesis
with GW observations beyond the level of null tests -- would be towards the study of rotating black holes, perhaps initially
within a slow-rotation approximation. These are all important issues that need to be addressed in future work.


\acknowledgements

We are grateful to Carlos Molina for providing us with black hole QNM data in dCS gravity.
We also wish to thank Caio F. B. Macedo for helpful feedback on our paper.
K.G. acknowledges networking support by the COST Actions GWverse CA16104 and PHAROS  CA16214.
H.O.S. acknowledges financial support through NASA Grants No.~NNX16AB98G and No.~80NSSC17M0041.


\appendix


\section{Photon geodesics in an arbitrary spherical metric}
\label{sec:geodesics}

This appendix provides a compact discussion of null geodesics in a general spherically symmetric spacetime of the form
(see also Ref.~\cite{Cardoso:2008bp} for a similar analysis)
\be
ds^2 = g_{tt} (r) dt^2 + g_{rr} (r) dr^2 + r^2 d\Omega^2.
\label{gspherical}
\ee
A photon's radial motion in this metric is described by
\be
(u^r )^2 = - \frac{1}{g_{rr}} \left ( \frac{1}{g_{tt}} + \frac{b^2}{r^2} \right ) \equiv V_r (r,b),
\label{pheom}
\ee
where $u^\mu = dx^\mu/d\lambda$ for some affine parameter $\lambda$ and $b= L/E$ is the orbital impact parameter.

For circular orbits of radius $r=\rph$, the `photon ring equation' materialises from the two conditions
$ V_r (\rph) = V_r^\p (\rph) =0$.
We find $ b^2_{\rm ph} = -\rph^2/g_{tt} (\rph) $ and
\be
\left ( \frac{g_{tt}^\p}{g_{tt}} \right )_{\rm ph} = \frac{2}{\rph}.
\label{rpheqsph2}
\ee
The resulting photon ring equation is independent of $g_{rr}$ and, of course, for the case of
the  Schwarzschild metric it is solved by $\rph =3 M$.

The angular frequency $\Omega_{\rm ph}$ at the photon ring is
\be
\Omega_{\rm ph} = \frac{u^\varphi}{u^t} =  \frac{1}{b_{\rm ph}} = \frac{\sqrt{-g_{tt}(\rph)}}{\rph}.
\label{Omph}
\ee
Another parameter of interest is the photon ring's Lyapunov exponent $\gamma_{\rm ph}$ which is a measure of the
rate of convergence/divergence of null rays in the ring's vicinity. A general expression for this parameter is given
by the stability calculation of Ref.~\cite{Cardoso:2008bp},
\be
\gamma^2_{\rm ph} = \frac{V_r^{\pp}}{2 (u^t)^2} = -\frac{1}{2} \left  [ r^2 g_{tt}  \left ( \frac{g_{tt}}{r^2} \right )^{\pp} \right ]_{\rm ph}.
\label{Lyap1}
\ee


\section{On the eikonal QNM-geodesic correspondence of the scalar wave equation}
\label{sec:qnmgeod}

In this appendix we consider the wave equation $\Box \Phi =0$ for a massless scalar field in the general spherical metric
(\ref{gspherical}) and demonstrate that the eikonal QNM is related to the spacetime's photon ring.
The standard decomposition in spherical coordinates,
\be
\Phi = \frac{1}{r} \psi (r) Y_\ell^m (\theta,\varphi) e^{-i\omega t},
\ee
leads to the following radial wave equation,
\be
\frac{d^2 \psi}{dx^2} + \left\{ \omega^2 + g_{tt} \left [\frac{\ell(\ell+1)}{r^2} + \frac{1}{2r g_{rr}} \left ( \frac{g_{tt}^\p}{g_{tt}}
- \frac{g_{rr}^\p}{g_{rr}} \right )  \right] \right\} \psi = 0,
\label{radscalar}
\ee
where the tortoise coordinate is defined as
\be
\frac{dx}{dr} = \left ( -\frac{g_{rr}}{ g_{tt}} \right )^{1/2}.
\ee
We can observe that in the eikonal limit $\ell \gg 1$ only the $g_{tt}$ component is relevant.
Furthermore, assuming a single-peak black hole-like potential, the eikonal analysis of Sec.~\ref{sec:wave_eik} applies,
and the real part of the fundamental mode is
\be
\omega_R \approx \ell\, \frac{\sqrt{-g_{tt} (r_0)}}{r_0}.
\label{omW}
\ee
The peak $r_0$ of the eikonal potential  $\tilde{U}(r) = g_{tt}(r)/r^2$ solves the equation $2 g_{tt} = r  g_{tt}^\p $
which coincides with the photon ring equation (\ref{rpheqsph2}) in a general spherical spacetime.
Then $r_0 = \rph$, and with the help of (\ref{Omph}), we can rewrite (\ref{omW}) in terms of the photon ring frequency as
\be
\omega_R \approx \ell \, \Omega_{\rm ph}.
\ee
A similar reasoning establishes a connection between the imaginary part $\omega_I$ [see Eq.~(\ref{omIeik1})] and the photon ring's
Lyapunov exponent. Using (\ref{Lyap1}) we first obtain
\be
\gamma^2_{\rm ph} =  -\frac{1}{2} \left ( g_{tt}^2 \frac{\tilde{U}^{\pp}}{\tilde{U}} \right )_{\rm ph}.
\label{Lyap2}
\ee
and then it is easy to show that
\be
\omega_I \approx -\frac{1}{2} | \gamma_{\rm ph}|,
\ee
We have thus found the same relations as the ones encountered in the eikonal study of QNMs of GR black holes~\cite{Ferrari:1984zz,Mashhoon:1985cya}.


\bibliography{biblio.bib}

\begin{thebibliography}{45}%
\makeatletter
\providecommand \@ifxundefined [1]{%
 \@ifx{#1\undefined}
}%
\providecommand \@ifnum [1]{%
 \ifnum #1\expandafter \@firstoftwo
 \else \expandafter \@secondoftwo
 \fi
}%
\providecommand \@ifx [1]{%
 \ifx #1\expandafter \@firstoftwo
 \else \expandafter \@secondoftwo
 \fi
}%
\providecommand \natexlab [1]{#1}%
\providecommand \enquote  [1]{``#1''}%
\providecommand \bibnamefont  [1]{#1}%
\providecommand \bibfnamefont [1]{#1}%
\providecommand \citenamefont [1]{#1}%
\providecommand \href@noop [0]{\@secondoftwo}%
\providecommand \href [0]{\begingroup \@sanitize@url \@href}%
\providecommand \@href[1]{\@@startlink{#1}\@@href}%
\providecommand \@@href[1]{\endgroup#1\@@endlink}%
\providecommand \@sanitize@url [0]{\catcode `\\12\catcode `\$12\catcode
  `\&12\catcode `\#12\catcode `\^12\catcode `\_12\catcode `\%12\relax}%
\providecommand \@@startlink[1]{}%
\providecommand \@@endlink[0]{}%
\providecommand \url  [0]{\begingroup\@sanitize@url \@url }%
\providecommand \@url [1]{\endgroup\@href {#1}{\urlprefix }}%
\providecommand \urlprefix  [0]{URL }%
\providecommand \Eprint [0]{\href }%
\providecommand \doibase [0]{http://dx.doi.org/}%
\providecommand \selectlanguage [0]{\@gobble}%
\providecommand \bibinfo  [0]{\@secondoftwo}%
\providecommand \bibfield  [0]{\@secondoftwo}%
\providecommand \translation [1]{[#1]}%
\providecommand \BibitemOpen [0]{}%
\providecommand \bibitemStop [0]{}%
\providecommand \bibitemNoStop [0]{.\EOS\space}%
\providecommand \EOS [0]{\spacefactor3000\relax}%
\providecommand \BibitemShut  [1]{\csname bibitem#1\endcsname}%
\let\auto@bib@innerbib\@empty
\bibitem [{\citenamefont {Abbott}\ \emph
  {et~al.}(2016{\natexlab{a}})\citenamefont {Abbott} \emph
  {et~al.}}]{Abbott:2016blz}%
  \BibitemOpen
  \bibfield  {author} {\bibinfo {author} {\bibfnamefont {B.~P.}\ \bibnamefont
  {Abbott}} \emph {et~al.} (\bibinfo {collaboration} {Virgo, LIGO
  Scientific}),\ }\href {\doibase 10.1103/PhysRevLett.116.061102} {\bibfield
  {journal} {\bibinfo  {journal} {Phys. Rev. Lett.}\ }\textbf {\bibinfo
  {volume} {116}},\ \bibinfo {pages} {061102} (\bibinfo {year}
  {2016}{\natexlab{a}})}\BibitemShut {NoStop}%
\bibitem [{\citenamefont {Abbott}\ \emph
  {et~al.}(2016{\natexlab{b}})\citenamefont {Abbott} \emph
  {et~al.}}]{TheLIGOScientific:2016pea}%
  \BibitemOpen
  \bibfield  {author} {\bibinfo {author} {\bibfnamefont {B.~P.}\ \bibnamefont
  {Abbott}} \emph {et~al.} (\bibinfo {collaboration} {Virgo, LIGO
  Scientific}),\ }\href {\doibase 10.1103/PhysRevX.6.041015} {\bibfield
  {journal} {\bibinfo  {journal} {Phys. Rev. X}\ }\textbf {\bibinfo {volume}
  {6}},\ \bibinfo {pages} {041015} (\bibinfo {year}
  {2016}{\natexlab{b}})}\BibitemShut {NoStop}%
\bibitem [{\citenamefont {Abbott}\ \emph
  {et~al.}(2016{\natexlab{c}})\citenamefont {Abbott} \emph
  {et~al.}}]{Abbott:2016nmj}%
  \BibitemOpen
  \bibfield  {author} {\bibinfo {author} {\bibfnamefont {B.~P.}\ \bibnamefont
  {Abbott}} \emph {et~al.} (\bibinfo {collaboration} {Virgo, LIGO
  Scientific}),\ }\href {\doibase 10.1103/PhysRevLett.116.241103} {\bibfield
  {journal} {\bibinfo  {journal} {Phys. Rev. Lett.}\ }\textbf {\bibinfo
  {volume} {116}},\ \bibinfo {pages} {241103} (\bibinfo {year}
  {2016}{\natexlab{c}})}\BibitemShut {NoStop}%
\bibitem [{\citenamefont {Abbott}\ \emph
  {et~al.}(2017{\natexlab{a}})\citenamefont {Abbott} \emph
  {et~al.}}]{Abbott:2017vtc}%
  \BibitemOpen
  \bibfield  {author} {\bibinfo {author} {\bibfnamefont {B.~P.}\ \bibnamefont
  {Abbott}} \emph {et~al.} (\bibinfo {collaboration} {VIRGO, LIGO
  Scientific}),\ }\href {\doibase 10.1103/PhysRevLett.118.221101} {\bibfield
  {journal} {\bibinfo  {journal} {Phys. Rev. Lett.}\ }\textbf {\bibinfo
  {volume} {118}},\ \bibinfo {pages} {221101} (\bibinfo {year}
  {2017}{\natexlab{a}})}\BibitemShut {NoStop}%
\bibitem [{\citenamefont {Abbott}\ \emph
  {et~al.}(2017{\natexlab{b}})\citenamefont {Abbott} \emph
  {et~al.}}]{GW170814}%
  \BibitemOpen
  \bibfield  {author} {\bibinfo {author} {\bibfnamefont {B.~P.}\ \bibnamefont
  {Abbott}} \emph {et~al.} (\bibinfo {collaboration} {Virgo, LIGO
  Scientific}),\ }\href {\doibase 10.1103/PhysRevLett.119.141101} {\bibfield
  {journal} {\bibinfo  {journal} {Phys. Rev. Lett.}\ }\textbf {\bibinfo
  {volume} {119}},\ \bibinfo {pages} {141101} (\bibinfo {year}
  {2017}{\natexlab{b}})}\BibitemShut {NoStop}%
\bibitem [{\citenamefont {Detweiler}(1980)}]{Detweiler:1980gk}%
  \BibitemOpen
  \bibfield  {author} {\bibinfo {author} {\bibfnamefont {S.~L.}\ \bibnamefont
  {Detweiler}},\ }\href {\doibase 10.1086/158109} {\bibfield  {journal}
  {\bibinfo  {journal} {Astrophys. J.}\ }\textbf {\bibinfo {volume} {239}},\
  \bibinfo {pages} {292} (\bibinfo {year} {1980})}\BibitemShut {NoStop}%
\bibitem [{\citenamefont {Dreyer}\ \emph {et~al.}(2004)\citenamefont {Dreyer},
  \citenamefont {Kelly}, \citenamefont {Krishnan}, \citenamefont {Finn},
  \citenamefont {Garrison},\ and\ \citenamefont
  {Lopez-Aleman}}]{Dreyer:2003bv}%
  \BibitemOpen
  \bibfield  {author} {\bibinfo {author} {\bibfnamefont {O.}~\bibnamefont
  {Dreyer}}, \bibinfo {author} {\bibfnamefont {B.~J.}\ \bibnamefont {Kelly}},
  \bibinfo {author} {\bibfnamefont {B.}~\bibnamefont {Krishnan}}, \bibinfo
  {author} {\bibfnamefont {L.~S.}\ \bibnamefont {Finn}}, \bibinfo {author}
  {\bibfnamefont {D.}~\bibnamefont {Garrison}}, \ and\ \bibinfo {author}
  {\bibfnamefont {R.}~\bibnamefont {Lopez-Aleman}},\ }\href {\doibase
  10.1088/0264-9381/21/4/003} {\bibfield  {journal} {\bibinfo  {journal}
  {Class. Quant. Grav.}\ }\textbf {\bibinfo {volume} {21}},\ \bibinfo {pages}
  {787} (\bibinfo {year} {2004})}\BibitemShut {NoStop}%
\bibitem [{\citenamefont {Berti}\ \emph {et~al.}(2006)\citenamefont {Berti},
  \citenamefont {Cardoso},\ and\ \citenamefont {Will}}]{Berti:2005ys}%
  \BibitemOpen
  \bibfield  {author} {\bibinfo {author} {\bibfnamefont {E.}~\bibnamefont
  {Berti}}, \bibinfo {author} {\bibfnamefont {V.}~\bibnamefont {Cardoso}}, \
  and\ \bibinfo {author} {\bibfnamefont {C.~M.}\ \bibnamefont {Will}},\ }\href
  {\doibase 10.1103/PhysRevD.73.064030} {\bibfield  {journal} {\bibinfo
  {journal} {Phys. Rev. D}\ }\textbf {\bibinfo {volume} {73}},\ \bibinfo
  {pages} {064030} (\bibinfo {year} {2006})}\BibitemShut {NoStop}%
\bibitem [{\citenamefont {Johannsen}\ and\ \citenamefont
  {Psaltis}(2011)}]{Johannsen:2011dh}%
  \BibitemOpen
  \bibfield  {author} {\bibinfo {author} {\bibfnamefont {T.}~\bibnamefont
  {Johannsen}}\ and\ \bibinfo {author} {\bibfnamefont {D.}~\bibnamefont
  {Psaltis}},\ }\href {\doibase 10.1103/PhysRevD.83.124015} {\bibfield
  {journal} {\bibinfo  {journal} {Phys. Rev. D}\ }\textbf {\bibinfo {volume}
  {83}},\ \bibinfo {pages} {124015} (\bibinfo {year} {2011})}\BibitemShut
  {NoStop}%
\bibitem [{\citenamefont {{Johannsen}}(2013)}]{Johannsen2013PhRvD}%
  \BibitemOpen
  \bibfield  {author} {\bibinfo {author} {\bibfnamefont {T.}~\bibnamefont
  {{Johannsen}}},\ }\href {\doibase 10.1103/PhysRevD.88.044002} {\bibfield
  {journal} {\bibinfo  {journal} {\prd}\ }\textbf {\bibinfo {volume} {88}},\
  \bibinfo {pages} {044002} (\bibinfo {year} {2013})}\BibitemShut {NoStop}%
\bibitem [{\citenamefont {Konoplya}\ \emph {et~al.}(2016)\citenamefont
  {Konoplya}, \citenamefont {Rezzolla},\ and\ \citenamefont
  {Zhidenko}}]{Konoplya:2016jvv}%
  \BibitemOpen
  \bibfield  {author} {\bibinfo {author} {\bibfnamefont {R.}~\bibnamefont
  {Konoplya}}, \bibinfo {author} {\bibfnamefont {L.}~\bibnamefont {Rezzolla}},
  \ and\ \bibinfo {author} {\bibfnamefont {A.}~\bibnamefont {Zhidenko}},\
  }\href {\doibase 10.1103/PhysRevD.93.064015} {\bibfield  {journal} {\bibinfo
  {journal} {Phys. Rev. D}\ }\textbf {\bibinfo {volume} {93}},\ \bibinfo
  {pages} {064015} (\bibinfo {year} {2016})}\BibitemShut {NoStop}%
\bibitem [{\citenamefont {Cardoso}\ \emph {et~al.}(2014)\citenamefont
  {Cardoso}, \citenamefont {Pani},\ and\ \citenamefont
  {Rico}}]{Cardoso:2014rha}%
  \BibitemOpen
  \bibfield  {author} {\bibinfo {author} {\bibfnamefont {V.}~\bibnamefont
  {Cardoso}}, \bibinfo {author} {\bibfnamefont {P.}~\bibnamefont {Pani}}, \
  and\ \bibinfo {author} {\bibfnamefont {J.}~\bibnamefont {Rico}},\ }\href
  {\doibase 10.1103/PhysRevD.89.064007} {\bibfield  {journal} {\bibinfo
  {journal} {Phys. Rev. D}\ }\textbf {\bibinfo {volume} {89}},\ \bibinfo
  {pages} {064007} (\bibinfo {year} {2014})}\BibitemShut {NoStop}%
\bibitem [{\citenamefont {Glampedakis}\ \emph {et~al.}(2017)\citenamefont
  {Glampedakis}, \citenamefont {Pappas}, \citenamefont {Silva},\ and\
  \citenamefont {Berti}}]{Glampedakis:2017}%
  \BibitemOpen
  \bibfield  {author} {\bibinfo {author} {\bibfnamefont {K.}~\bibnamefont
  {Glampedakis}}, \bibinfo {author} {\bibfnamefont {G.}~\bibnamefont {Pappas}},
  \bibinfo {author} {\bibfnamefont {H.~O.}\ \bibnamefont {Silva}}, \ and\
  \bibinfo {author} {\bibfnamefont {E.}~\bibnamefont {Berti}},\ }\href
  {\doibase 10.1103/PhysRevD.96.064054} {\bibfield  {journal} {\bibinfo
  {journal} {Phys. Rev. D}\ }\textbf {\bibinfo {volume} {96}},\ \bibinfo
  {pages} {064054} (\bibinfo {year} {2017})}\BibitemShut {NoStop}%
\bibitem [{\citenamefont {Cardoso}\ \emph {et~al.}(2019)\citenamefont
  {Cardoso}, \citenamefont {Kimura}, \citenamefont {Maselli}, \citenamefont
  {Berti}, \citenamefont {Macedo},\ and\ \citenamefont
  {McManus}}]{Cardoso_etal2019}%
  \BibitemOpen
  \bibfield  {author} {\bibinfo {author} {\bibfnamefont {V.}~\bibnamefont
  {Cardoso}}, \bibinfo {author} {\bibfnamefont {M.}~\bibnamefont {Kimura}},
  \bibinfo {author} {\bibfnamefont {A.}~\bibnamefont {Maselli}}, \bibinfo
  {author} {\bibfnamefont {E.}~\bibnamefont {Berti}}, \bibinfo {author}
  {\bibfnamefont {C.~F.~B.}\ \bibnamefont {Macedo}}, \ and\ \bibinfo {author}
  {\bibfnamefont {R.}~\bibnamefont {McManus}},\ }\href@noop {} {\  (\bibinfo
  {year} {2019})},\ \Eprint {http://arxiv.org/abs/1901.01265} {arXiv:1901.01265
  [gr-qc]} \BibitemShut {NoStop}%
\bibitem [{\citenamefont {McManus}\ \emph {et~al.}(2019)\citenamefont
  {McManus}, \citenamefont {Berti}, \citenamefont {Macedo}, \citenamefont
  {Kimura}, \citenamefont {Maselli},\ and\ \citenamefont
  {Cardoso}}]{McManus_etal2019}%
  \BibitemOpen
  \bibfield  {author} {\bibinfo {author} {\bibfnamefont {R.}~\bibnamefont
  {McManus}}, \bibinfo {author} {\bibfnamefont {E.}~\bibnamefont {Berti}},
  \bibinfo {author} {\bibfnamefont {C.~F.~B.}\ \bibnamefont {Macedo}}, \bibinfo
  {author} {\bibfnamefont {M.}~\bibnamefont {Kimura}}, \bibinfo {author}
  {\bibfnamefont {A.}~\bibnamefont {Maselli}}, \ and\ \bibinfo {author}
  {\bibfnamefont {V.}~\bibnamefont {Cardoso}},\ }\href@noop {} {\  (\bibinfo
  {year} {2019})},\ \Eprint {http://arxiv.org/abs/1906.05155} {arXiv:1906.05155
  [gr-qc]} \BibitemShut {NoStop}%
\bibitem [{\citenamefont {Molina}\ \emph {et~al.}(2010)\citenamefont {Molina},
  \citenamefont {Pani}, \citenamefont {Cardoso},\ and\ \citenamefont
  {Gualtieri}}]{molina2010}%
  \BibitemOpen
  \bibfield  {author} {\bibinfo {author} {\bibfnamefont {C.}~\bibnamefont
  {Molina}}, \bibinfo {author} {\bibfnamefont {P.}~\bibnamefont {Pani}},
  \bibinfo {author} {\bibfnamefont {V.}~\bibnamefont {Cardoso}}, \ and\
  \bibinfo {author} {\bibfnamefont {L.}~\bibnamefont {Gualtieri}},\ }\href
  {\doibase https://doi.org/10.1103/PhysRevD.81.124021} {\bibfield  {journal}
  {\bibinfo  {journal} {Phys. Rev. D}\ }\textbf {\bibinfo {volume} {81}},\
  \bibinfo {pages} {124021} (\bibinfo {year} {2010})}\BibitemShut {NoStop}%
\bibitem [{\citenamefont {Kobayashi}\ \emph {et~al.}(2012)\citenamefont
  {Kobayashi}, \citenamefont {Motohashi},\ and\ \citenamefont
  {Suyama}}]{Kobayashi:2012kh}%
  \BibitemOpen
  \bibfield  {author} {\bibinfo {author} {\bibfnamefont {T.}~\bibnamefont
  {Kobayashi}}, \bibinfo {author} {\bibfnamefont {H.}~\bibnamefont
  {Motohashi}}, \ and\ \bibinfo {author} {\bibfnamefont {T.}~\bibnamefont
  {Suyama}},\ }\href {\doibase 10.1103/PhysRevD.85.084025} {\bibfield
  {journal} {\bibinfo  {journal} {Phys. Rev. D}\ }\textbf {\bibinfo {volume}
  {85}},\ \bibinfo {pages} {084025} (\bibinfo {year} {2012})}\BibitemShut
  {NoStop}%
\bibitem [{\citenamefont {Kobayashi}\ \emph {et~al.}(2014)\citenamefont
  {Kobayashi}, \citenamefont {Motohashi},\ and\ \citenamefont
  {Suyama}}]{Kobayashi:2014wsa}%
  \BibitemOpen
  \bibfield  {author} {\bibinfo {author} {\bibfnamefont {T.}~\bibnamefont
  {Kobayashi}}, \bibinfo {author} {\bibfnamefont {H.}~\bibnamefont
  {Motohashi}}, \ and\ \bibinfo {author} {\bibfnamefont {T.}~\bibnamefont
  {Suyama}},\ }\href {\doibase 10.1103/PhysRevD.89.084042} {\bibfield
  {journal} {\bibinfo  {journal} {Phys. Rev. D}\ }\textbf {\bibinfo {volume}
  {89}},\ \bibinfo {pages} {084042} (\bibinfo {year} {2014})}\BibitemShut
  {NoStop}%
\bibitem [{\citenamefont {Bl{\'a}zquez-Salcedo}\ \emph
  {et~al.}(2016)\citenamefont {Bl{\'a}zquez-Salcedo}, \citenamefont {Macedo},
  \citenamefont {Cardoso}, \citenamefont {Ferrari}, \citenamefont {Gualtieri},
  \citenamefont {Khoo}, \citenamefont {Kunz},\ and\ \citenamefont
  {Pani}}]{Blazquez-Salcedo:2016enn}%
  \BibitemOpen
  \bibfield  {author} {\bibinfo {author} {\bibfnamefont {J.~L.}\ \bibnamefont
  {Bl{\'a}zquez-Salcedo}}, \bibinfo {author} {\bibfnamefont {C.~F.~B.}\
  \bibnamefont {Macedo}}, \bibinfo {author} {\bibfnamefont {V.}~\bibnamefont
  {Cardoso}}, \bibinfo {author} {\bibfnamefont {V.}~\bibnamefont {Ferrari}},
  \bibinfo {author} {\bibfnamefont {L.}~\bibnamefont {Gualtieri}}, \bibinfo
  {author} {\bibfnamefont {F.~S.}\ \bibnamefont {Khoo}}, \bibinfo {author}
  {\bibfnamefont {J.}~\bibnamefont {Kunz}}, \ and\ \bibinfo {author}
  {\bibfnamefont {P.}~\bibnamefont {Pani}},\ }\href {\doibase
  10.1103/PhysRevD.94.104024} {\bibfield  {journal} {\bibinfo  {journal} {Phys.
  Rev. D}\ }\textbf {\bibinfo {volume} {94}},\ \bibinfo {pages} {104024}
  (\bibinfo {year} {2016})}\BibitemShut {NoStop}%
\bibitem [{\citenamefont {Bl{\'a}zquez-Salcedo}\ \emph
  {et~al.}(2017)\citenamefont {Bl{\'a}zquez-Salcedo}, \citenamefont {Khoo},\
  and\ \citenamefont {Kunz}}]{Blazquez-Salcedo:2017txk}%
  \BibitemOpen
  \bibfield  {author} {\bibinfo {author} {\bibfnamefont {J.~L.}\ \bibnamefont
  {Bl{\'a}zquez-Salcedo}}, \bibinfo {author} {\bibfnamefont {F.~S.}\
  \bibnamefont {Khoo}}, \ and\ \bibinfo {author} {\bibfnamefont
  {J.}~\bibnamefont {Kunz}},\ }\href {\doibase 10.1103/PhysRevD.96.064008}
  {\bibfield  {journal} {\bibinfo  {journal} {Phys. Rev. D}\ }\textbf {\bibinfo
  {volume} {96}},\ \bibinfo {pages} {064008} (\bibinfo {year}
  {2017})}\BibitemShut {NoStop}%
\bibitem [{\citenamefont {Brito}\ and\ \citenamefont
  {Pacilio}(2018)}]{Brito:2018hjh}%
  \BibitemOpen
  \bibfield  {author} {\bibinfo {author} {\bibfnamefont {R.}~\bibnamefont
  {Brito}}\ and\ \bibinfo {author} {\bibfnamefont {C.}~\bibnamefont
  {Pacilio}},\ }\href {\doibase 10.1103/PhysRevD.98.104042} {\bibfield
  {journal} {\bibinfo  {journal} {Phys. Rev. D}\ }\textbf {\bibinfo {volume}
  {98}},\ \bibinfo {pages} {104042} (\bibinfo {year} {2018})}\BibitemShut
  {NoStop}%
\bibitem [{\citenamefont {Tattersall}\ and\ \citenamefont
  {Ferreira}(2018)}]{Tattersall_2018a}%
  \BibitemOpen
  \bibfield  {author} {\bibinfo {author} {\bibfnamefont {O.~J.}\ \bibnamefont
  {Tattersall}}\ and\ \bibinfo {author} {\bibfnamefont {P.~G.}\ \bibnamefont
  {Ferreira}},\ }\href {\doibase 10.1103/PhysRevD.97.104047} {\bibfield
  {journal} {\bibinfo  {journal} {Phys. Rev. D}\ }\textbf {\bibinfo {volume}
  {97}},\ \bibinfo {pages} {104047} (\bibinfo {year} {2018})}\BibitemShut
  {NoStop}%
\bibitem [{\citenamefont {Tattersall}\ \emph {et~al.}(2018)\citenamefont
  {Tattersall}, \citenamefont {Ferreira},\ and\ \citenamefont
  {Lagos}}]{Tattersall_etal2018}%
  \BibitemOpen
  \bibfield  {author} {\bibinfo {author} {\bibfnamefont {O.~J.}\ \bibnamefont
  {Tattersall}}, \bibinfo {author} {\bibfnamefont {P.~G.}\ \bibnamefont
  {Ferreira}}, \ and\ \bibinfo {author} {\bibfnamefont {M.}~\bibnamefont
  {Lagos}},\ }\href {\doibase 10.1103/PhysRevD.97.044021} {\bibfield  {journal}
  {\bibinfo  {journal} {Phys. Rev. D}\ }\textbf {\bibinfo {volume} {97}},\
  \bibinfo {pages} {044021} (\bibinfo {year} {2018})}\BibitemShut {NoStop}%
\bibitem [{\citenamefont {Berti}\ \emph {et~al.}(2015)\citenamefont {Berti}
  \emph {et~al.}}]{Bertietal2015}%
  \BibitemOpen
  \bibfield  {author} {\bibinfo {author} {\bibfnamefont {E.}~\bibnamefont
  {Berti}} \emph {et~al.},\ }\href {\doibase 10.1088/0264-9381/32/24/243001}
  {\bibfield  {journal} {\bibinfo  {journal} {Class. Quant. Grav.}\ }\textbf
  {\bibinfo {volume} {32}},\ \bibinfo {pages} {243001} (\bibinfo {year}
  {2015})}\BibitemShut {NoStop}%
\bibitem [{\citenamefont {Heisenberg}\ \emph {et~al.}(2017)\citenamefont
  {Heisenberg}, \citenamefont {Kase}, \citenamefont {Minamitsuji},\ and\
  \citenamefont {Tsujikawa}}]{Heisenberg:2017hwb}%
  \BibitemOpen
  \bibfield  {author} {\bibinfo {author} {\bibfnamefont {L.}~\bibnamefont
  {Heisenberg}}, \bibinfo {author} {\bibfnamefont {R.}~\bibnamefont {Kase}},
  \bibinfo {author} {\bibfnamefont {M.}~\bibnamefont {Minamitsuji}}, \ and\
  \bibinfo {author} {\bibfnamefont {S.}~\bibnamefont {Tsujikawa}},\ }\href
  {\doibase 10.1088/1475-7516/2017/08/024} {\bibfield  {journal} {\bibinfo
  {journal} {JCAP}\ }\textbf {\bibinfo {volume} {1708}},\ \bibinfo {pages}
  {024} (\bibinfo {year} {2017})}\BibitemShut {NoStop}%
\bibitem [{\citenamefont {Press}(1971)}]{Press:1971wr}%
  \BibitemOpen
  \bibfield  {author} {\bibinfo {author} {\bibfnamefont {W.~H.}\ \bibnamefont
  {Press}},\ }\href {\doibase 10.1086/180849} {\bibfield  {journal} {\bibinfo
  {journal} {Astrophys. J.}\ }\textbf {\bibinfo {volume} {170}},\ \bibinfo
  {pages} {L105} (\bibinfo {year} {1971})}\BibitemShut {NoStop}%
\bibitem [{\citenamefont {{Goebel}}(1972)}]{Goebel:1972}%
  \BibitemOpen
  \bibfield  {author} {\bibinfo {author} {\bibfnamefont {C.~J.}\ \bibnamefont
  {{Goebel}}},\ }\href {\doibase 10.1086/180898} {\bibfield  {journal}
  {\bibinfo  {journal} {Astrophys. J.}\ }\textbf {\bibinfo {volume} {172}},\
  \bibinfo {pages} {L95} (\bibinfo {year} {1972})}\BibitemShut {NoStop}%
\bibitem [{\citenamefont {Kokkotas}\ and\ \citenamefont
  {Schmidt}(1999)}]{Kokkotas:1999bd}%
  \BibitemOpen
  \bibfield  {author} {\bibinfo {author} {\bibfnamefont {K.~D.}\ \bibnamefont
  {Kokkotas}}\ and\ \bibinfo {author} {\bibfnamefont {B.~G.}\ \bibnamefont
  {Schmidt}},\ }\href {\doibase 10.12942/lrr-1999-2} {\bibfield  {journal}
  {\bibinfo  {journal} {Living Rev. Rel.}\ }\textbf {\bibinfo {volume} {2}},\
  \bibinfo {pages} {2} (\bibinfo {year} {1999})}\BibitemShut {NoStop}%
\bibitem [{\citenamefont {Berti}\ \emph {et~al.}(2009)\citenamefont {Berti},
  \citenamefont {Cardoso},\ and\ \citenamefont {Starinets}}]{Berti:2009kk}%
  \BibitemOpen
  \bibfield  {author} {\bibinfo {author} {\bibfnamefont {E.}~\bibnamefont
  {Berti}}, \bibinfo {author} {\bibfnamefont {V.}~\bibnamefont {Cardoso}}, \
  and\ \bibinfo {author} {\bibfnamefont {A.~O.}\ \bibnamefont {Starinets}},\
  }\href {\doibase 10.1088/0264-9381/26/16/163001} {\bibfield  {journal}
  {\bibinfo  {journal} {Class. Quant. Grav.}\ }\textbf {\bibinfo {volume}
  {26}},\ \bibinfo {pages} {163001} (\bibinfo {year} {2009})}\BibitemShut
  {NoStop}%
\bibitem [{\citenamefont {Ferrari}\ and\ \citenamefont
  {Mashhoon}(1984)}]{Ferrari:1984zz}%
  \BibitemOpen
  \bibfield  {author} {\bibinfo {author} {\bibfnamefont {V.}~\bibnamefont
  {Ferrari}}\ and\ \bibinfo {author} {\bibfnamefont {B.}~\bibnamefont
  {Mashhoon}},\ }\href {\doibase 10.1103/PhysRevD.30.295} {\bibfield  {journal}
  {\bibinfo  {journal} {Phys. Rev. D}\ }\textbf {\bibinfo {volume} {30}},\
  \bibinfo {pages} {295} (\bibinfo {year} {1984})}\BibitemShut {NoStop}%
\bibitem [{\citenamefont {Mashhoon}(1985)}]{Mashhoon:1985cya}%
  \BibitemOpen
  \bibfield  {author} {\bibinfo {author} {\bibfnamefont {B.}~\bibnamefont
  {Mashhoon}},\ }\href {\doibase 10.1103/PhysRevD.31.290} {\bibfield  {journal}
  {\bibinfo  {journal} {Phys. Rev. D}\ }\textbf {\bibinfo {volume} {31}},\
  \bibinfo {pages} {290} (\bibinfo {year} {1985})}\BibitemShut {NoStop}%
\bibitem [{\citenamefont {Cardoso}\ \emph {et~al.}(2009)\citenamefont
  {Cardoso}, \citenamefont {Miranda}, \citenamefont {Berti}, \citenamefont
  {Witek},\ and\ \citenamefont {Zanchin}}]{Cardoso:2008bp}%
  \BibitemOpen
  \bibfield  {author} {\bibinfo {author} {\bibfnamefont {V.}~\bibnamefont
  {Cardoso}}, \bibinfo {author} {\bibfnamefont {A.~S.}\ \bibnamefont
  {Miranda}}, \bibinfo {author} {\bibfnamefont {E.}~\bibnamefont {Berti}},
  \bibinfo {author} {\bibfnamefont {H.}~\bibnamefont {Witek}}, \ and\ \bibinfo
  {author} {\bibfnamefont {V.~T.}\ \bibnamefont {Zanchin}},\ }\href {\doibase
  10.1103/PhysRevD.79.064016} {\bibfield  {journal} {\bibinfo  {journal} {Phys.
  Rev. D}\ }\textbf {\bibinfo {volume} {79}},\ \bibinfo {pages} {064016}
  (\bibinfo {year} {2009})}\BibitemShut {NoStop}%
\bibitem [{\citenamefont {Dolan}(2010)}]{Dolan:2010wr}%
  \BibitemOpen
  \bibfield  {author} {\bibinfo {author} {\bibfnamefont {S.~R.}\ \bibnamefont
  {Dolan}},\ }\href {\doibase 10.1103/PhysRevD.82.104003} {\bibfield  {journal}
  {\bibinfo  {journal} {Phys. Rev. D}\ }\textbf {\bibinfo {volume} {82}},\
  \bibinfo {pages} {104003} (\bibinfo {year} {2010})}\BibitemShut {NoStop}%
\bibitem [{\citenamefont {Yang}\ \emph {et~al.}(2012)\citenamefont {Yang},
  \citenamefont {Nichols}, \citenamefont {Zhang}, \citenamefont {Zimmerman},
  \citenamefont {Zhang},\ and\ \citenamefont {Chen}}]{Yang:2012he}%
  \BibitemOpen
  \bibfield  {author} {\bibinfo {author} {\bibfnamefont {H.}~\bibnamefont
  {Yang}}, \bibinfo {author} {\bibfnamefont {D.~A.}\ \bibnamefont {Nichols}},
  \bibinfo {author} {\bibfnamefont {F.}~\bibnamefont {Zhang}}, \bibinfo
  {author} {\bibfnamefont {A.}~\bibnamefont {Zimmerman}}, \bibinfo {author}
  {\bibfnamefont {Z.}~\bibnamefont {Zhang}}, \ and\ \bibinfo {author}
  {\bibfnamefont {Y.}~\bibnamefont {Chen}},\ }\href {\doibase
  10.1103/PhysRevD.86.104006} {\bibfield  {journal} {\bibinfo  {journal} {Phys.
  Rev. D}\ }\textbf {\bibinfo {volume} {86}},\ \bibinfo {pages} {104006}
  (\bibinfo {year} {2012})}\BibitemShut {NoStop}%
\bibitem [{\citenamefont {Khanna}\ and\ \citenamefont
  {Price}(2017)}]{Khanna:2016yow}%
  \BibitemOpen
  \bibfield  {author} {\bibinfo {author} {\bibfnamefont {G.}~\bibnamefont
  {Khanna}}\ and\ \bibinfo {author} {\bibfnamefont {R.~H.}\ \bibnamefont
  {Price}},\ }\href {\doibase 10.1103/PhysRevD.95.081501} {\bibfield  {journal}
  {\bibinfo  {journal} {Phys. Rev. D}\ }\textbf {\bibinfo {volume} {95}},\
  \bibinfo {pages} {081501} (\bibinfo {year} {2017})}\BibitemShut {NoStop}%
\bibitem [{\citenamefont {Konoplya}\ and\ \citenamefont
  {Stuchl{\'\i}k}(2017)}]{Konoplya:2017wot}%
  \BibitemOpen
  \bibfield  {author} {\bibinfo {author} {\bibfnamefont {R.~A.}\ \bibnamefont
  {Konoplya}}\ and\ \bibinfo {author} {\bibfnamefont {Z.}~\bibnamefont
  {Stuchl{\'\i}k}},\ }\href {\doibase 10.1016/j.physletb.2017.06.015}
  {\bibfield  {journal} {\bibinfo  {journal} {Phys. Lett. B}\ }\textbf
  {\bibinfo {volume} {771}},\ \bibinfo {pages} {597} (\bibinfo {year}
  {2017})}\BibitemShut {NoStop}%
\bibitem [{\citenamefont {Berti}(2014)}]{Berti:2014bla}%
  \BibitemOpen
  \bibfield  {author} {\bibinfo {author} {\bibfnamefont {E.}~\bibnamefont
  {Berti}}\ }(\bibinfo {year} {2014})\ \Eprint {http://arxiv.org/abs/1410.4481}
  {arXiv:1410.4481 [gr-qc]} \BibitemShut {NoStop}%
\bibitem [{\citenamefont {Schutz}\ and\ \citenamefont
  {Will}(1985)}]{SchutzWill}%
  \BibitemOpen
  \bibfield  {author} {\bibinfo {author} {\bibfnamefont {B.~F.}\ \bibnamefont
  {Schutz}}\ and\ \bibinfo {author} {\bibfnamefont {C.~M.}\ \bibnamefont
  {Will}},\ }\href {\doibase https://doi.org/10.1086/184453} {\bibfield
  {journal} {\bibinfo  {journal} {Astrophys. J.}\ }\textbf {\bibinfo {volume}
  {291}},\ \bibinfo {pages} {L33} (\bibinfo {year} {1985})}\BibitemShut
  {NoStop}%
\bibitem [{\citenamefont {Chandrasekhar}(2002)}]{ChandraBook}%
  \BibitemOpen
  \bibfield  {author} {\bibinfo {author} {\bibfnamefont {S.}~\bibnamefont
  {Chandrasekhar}},\ }\href@noop {} {\emph {\bibinfo {title} {{The mathematical
  theory of black holes}}}}\ (\bibinfo  {publisher} {Oxford Univ. Press},\
  \bibinfo {address} {Oxford},\ \bibinfo {year} {2002})\BibitemShut {NoStop}%
\bibitem [{\citenamefont {Jackiw}\ and\ \citenamefont {Pi}(2003)}]{jackiw2003}%
  \BibitemOpen
  \bibfield  {author} {\bibinfo {author} {\bibfnamefont {R.}~\bibnamefont
  {Jackiw}}\ and\ \bibinfo {author} {\bibfnamefont {S.~Y.}\ \bibnamefont
  {Pi}},\ }\href {\doibase 10.1103/PhysRevD.68.104012} {\bibfield  {journal}
  {\bibinfo  {journal} {Phys. Rev. D}\ }\textbf {\bibinfo {volume} {68}},\
  \bibinfo {pages} {104012} (\bibinfo {year} {2003})}\BibitemShut {NoStop}%
\bibitem [{\citenamefont {Alexander}\ and\ \citenamefont
  {Yunes}(2009)}]{alexander2009}%
  \BibitemOpen
  \bibfield  {author} {\bibinfo {author} {\bibfnamefont {S.}~\bibnamefont
  {Alexander}}\ and\ \bibinfo {author} {\bibfnamefont {N.}~\bibnamefont
  {Yunes}},\ }\href {\doibase https://doi.org/10.1016/j.physrep.2009.07.002}
  {\bibfield  {journal} {\bibinfo  {journal} {Phys. Rept.}\ }\textbf {\bibinfo
  {volume} {480}},\ \bibinfo {pages} {1} (\bibinfo {year} {2009})}\BibitemShut
  {NoStop}%
\bibitem [{\citenamefont {Yunes}\ and\ \citenamefont
  {Pretorius}(2009)}]{yunes2009}%
  \BibitemOpen
  \bibfield  {author} {\bibinfo {author} {\bibfnamefont {N.}~\bibnamefont
  {Yunes}}\ and\ \bibinfo {author} {\bibfnamefont {F.}~\bibnamefont
  {Pretorius}},\ }\href {\doibase 10.1103/PhysRevD.79.084043} {\bibfield
  {journal} {\bibinfo  {journal} {Phys. Rev. D}\ }\textbf {\bibinfo {volume}
  {79}},\ \bibinfo {pages} {084043} (\bibinfo {year} {2009})}\BibitemShut
  {NoStop}%
\bibitem [{\citenamefont {Yunes}\ and\ \citenamefont
  {Sopuerta}(2008)}]{Yunes2008}%
  \BibitemOpen
  \bibfield  {author} {\bibinfo {author} {\bibfnamefont {N.}~\bibnamefont
  {Yunes}}\ and\ \bibinfo {author} {\bibfnamefont {C.~F.}\ \bibnamefont
  {Sopuerta}},\ }\href {\doibase 10.1103/PhysRevD.77.064007} {\bibfield
  {journal} {\bibinfo  {journal} {Phys. Rev. D}\ }\textbf {\bibinfo {volume}
  {77}},\ \bibinfo {pages} {064007} (\bibinfo {year} {2008})}\BibitemShut
  {NoStop}%
\bibitem [{\citenamefont {Cardoso}\ and\ \citenamefont
  {Gualtieri}(2009)}]{Cardoso:2009pk}%
  \BibitemOpen
  \bibfield  {author} {\bibinfo {author} {\bibfnamefont {V.}~\bibnamefont
  {Cardoso}}\ and\ \bibinfo {author} {\bibfnamefont {L.}~\bibnamefont
  {Gualtieri}},\ }\href {\doibase 10.1103/PhysRevD.81.089903,
  10.1103/PhysRevD.80.064008} {\bibfield  {journal} {\bibinfo  {journal} {Phys.
  Rev. D}\ }\textbf {\bibinfo {volume} {80}},\ \bibinfo {pages} {064008}
  (\bibinfo {year} {2009})},\ \bibinfo {note} {[Erratum: Phys. Rev. D 81,089903
  (2010)]}\BibitemShut {NoStop}%
\bibitem [{\citenamefont {Molina}()}]{molina_private}%
  \BibitemOpen
  \bibfield  {author} {\bibinfo {author} {\bibfnamefont {C.}~\bibnamefont
  {Molina}},\ }\href@noop {} {}\bibinfo {note} {{private
  communication}}\BibitemShut {NoStop}%
\end{thebibliography}%


\end{document}